\documentclass[a4paper,fleqn]{cas-sc}

\usepackage[authoryear]{natbib}
\usepackage{graphicx}
\usepackage{bm}
\usepackage{textcomp}
\usepackage{threeparttable}

\def\tsc#1{\csdef{#1}{\textsc{\lowercase{#1}}\xspace}}
\tsc{WGM}
\tsc{QE}
\begin{document}
\let\WriteBookmarks\relax
\def\floatpagepagefraction{1}
\def\textpagefraction{.001}

\shorttitle{Adaptive Sampling with PIXL on the Mars {\it Perseverance} Rover}    

\shortauthors{P.R. Lawson, T.V. Kizovski, M.M. Tice, et al.}

\title [mode = title]{Adaptive Sampling with PIXL on the Mars {\it Perseverance} Rover}

%\author[<aff no>]{<author name>}[role=Researcher]
\author[1]{Peter R. Lawson}[orcid=0009-0004-6928-1565]
\cormark[1]
\ead{prlawson37@gmail.com}
\credit{Conceptualization, Methodology, Software, Validation, Formal analysis, Investigation, Writing - Original Draft, Writing - Review \& Editing}
\affiliation[1]{organization={Retired - Jet Propulsion Laboratory, California Institute of Technology},
            addressline={4800 Oak Grove Drive}, 
            city={Pasadena},
            postcode={91109}, 
            state={CA},
            country={USA}}
\cortext[cor1]{Corresponding author.}

\author[2]{Tanya V. Kizovski}
\credit{Methodology, Validation, Formal analysis, Writing - Original Draft, Writing - Review \& Editing}
\affiliation[2]{organization={Brock University},
            addressline={1812 Sir Isaac Brock Way},
            city={St. Catherines},
            postcode={L2S 3A1},
            state={ON},
            country={Canada}}

\author[3]{Michael M. Tice}
\credit{Conceptualization, Methodology, Validation, Formal analysis, Writing - Review \& Editing}
\affiliation[3]{organization={Texas A\&M University},
            addressline={3115 TAMU}, 
            city={College Station},
            postcode={77843-3115}, 
            state={TX},
            country={USA}}

\author[4]{Benton C. {Clark III}}
\credit{Conceptualization, Validation, Formal analysis, Writing - Review \& Editing}
\affiliation[4]{organization={Space Science Institute},
            addressline={4765 Walnut St, Suite B},
            city={Boulder},
            postcode={80301},
            state={CO},
            country={USA}}

\author[5]{Scott J. VanBommel}
\credit{Formal analysis, Writing - Review \& Editing}
\affiliation[5]{organization={Washington University in St. Louis},
            addressline={1 Brookings Drive},
            city={St. Louis},
            postcode={63130},
            state={MO},
            country={USA}}

\author[6]{David R. Thompson}
\credit{Conceptualization, Methodology, Software, Writing - Original Draft}
\affiliation[6]{organization={Jet Propulsion Laboratory, California Institute of Technology},
            addressline={4800 Oak Grove Drive}, 
            city={Pasadena},
            postcode={91109}, 
            state={CA},
            country={USA}}

\author[6]{Lawrence A. Wade}
\credit{Conceptualization, Methodology}

\author[1]{Robert W. Denise}
\credit{Software}

\author[6]{Christopher M. Heirwegh}
\credit{Software, Data Curation, Writing - Review \& Editing}

\author[7]{W. Timothy Elam}
\credit{Methodology, Software, Data Curation}
\affiliation[7]{organization={University of Washington},
            addressline={1013 NE 40th St.},
            city={Seattle},
            postcode={98105},
            state={WA},
            country={USA}}

\author[2]{Mariek E. Schmidt}
\credit{Formal analysis, Writing - Review \& Editing}

\author[6]{Yang Liu}
\credit{Conceptualization, Methodology, Writing - Review \& Editing}

\author[6]{Abigail C. Allwood}
\credit{Conceptualization, Funding acquisition}

\author[6]{Martin S. Gilbert}
\credit{Software}

\author[6]{Benjamin J. Bornstein}
\credit{Software}

\begin{abstract}
Planetary rovers can use onboard data analysis to adapt their measurement plan on the fly, improving the science value of data collected between commands from Earth.
This paper describes the implementation of an adaptive sampling algorithm used by PIXL, 
the X-ray fluorescence spectrometer of the Mars 2020 {\it Perseverance} rover. PIXL is deployed using 
the rover arm to measure X-ray spectra of rocks 
with a scan density of several thousand points over an area of typically 5 x 7 mm. The adaptive sampling algorithm is 
programmed to recognize points of interest and to increase the signal-to-noise ratio at those locations by performing longer
integrations. Two approaches are used to formulate the sampling rules based on past quantification data: 1) Expressions that isolate 
particular regions within a ternary compositional diagram, and 2) Machine learning rules that threshold for a high weight percent of particular
compounds.  The design of the rulesets are outlined and the performance of the algorithm is quantified using 
measurements from the surface of Mars.
To our knowledge, PIXL's adaptive sampling represents the first 
autonomous decision-making based on real-time compositional analysis by a spacecraft on the surface of another planet.
\end{abstract}

\begin{highlights}
\item First autonomous decision-making is demonstrated based on real-time compositional analysis. 
\item Adaptive sampling rules are created using X-ray spectra of rocks observed by the Mars {\it Perseverance} rover. 
\item Rules are formulated through machine learning and the analysis of ternary compositional diagrams.
\item Ground-based simulations exactly reproduce the behavior observed on Mars.
\item Long dwells are reliably triggered for carbonates, phosphate minerals, and spinels.
\end{highlights}


\maketitle

\section{Introduction}\label{introduction}

As successive generations of Mars rovers have become more capable, and their operational objectives more ambitious, they have made increasing use of spacecraft autonomy to achieve their science objectives.  Operators have limited communications opportunities with the remote spacecraft, so autonomy can help maximize the productivity of rover activity during the finite mission lifetime \citep{candela2017,wettergreen2014,thompson2011}.  The Mars 2020 {\it Perseverance} rover exemplifies this trend.  This rover must travel long distances across many geologic units to obtain a diverse collection of samples for return to Earth, traveling quickly between sample sites.  Its rapid exploration is made possible by improvements in autonomous navigation  \citep{farley2020,verma_2023}.  Upon reaching a site of interest, the rover has only limited resources 
for data collection to establish the geologic context.  Here autonomy can also play a role, by adapting the data collection activities on the fly and improving the science yield of each command cycle.  

This {\it science autonomy} has been deployed to Mars several times before {\it Perseverance}.  The Mars Exploration Rovers used onboard image analysis to capture images of dust devils on the Mars surface \citep{castano2008}.  The Opportunity rover used onboard rock detection to identify targets of interest for followup imagery during long drives \citep{estlin2012}. This capability was later developed into an onboard targeting procedure used by the Curiosity rover to point its ChemCam instrument \citep{francis2017}.  
All of these algorithms followed a similar pattern in which the rover identified opportunistic science targets in image data and then prioritized key features for followup data collection.
Another notable example of science autonomy is the 
range-finding algorithm used by Curiosity's Alpha Particle X-ray Spectrometer (APXS) 
 to improve placement proximity, particularly when the target surface is unconsolidated and/or potentially poses a deployment hazard
\citep{vanbommel_2019}.

This paper extends the concept of adaptive rover data collection to compositional spectroscopic analysis.  
Specifically, we describe the adaptive data collection algorithms used by the Planetary Instrument for X-ray Lithochemistry (PIXL), a rastering X-ray fluorescence spectrometer onboard the Mars {\it Perseverance} rover \citep{allwood2020}.  PIXL's onboard adaptive sampling procedures perform a simple real-time analysis of acquired spectra, optimizing its limited acquisition time budget to focus on any features of scientific interest that are discovered during data collection.  This gathers higher fidelity data from targets of interest that were not anticipated, and which --- due to the challenges of instrument placement or the need to move the rover to another location --- may not be measured again.
At the time of this writing, PIXL's adaptive sampling capability has been operating on Mars for over 951 sols (martian days), with over 
% It was first used on Sol 186. 
% May 1, 2024 is sol 1137. 1137 - 186 = 951. 
% Adaptive sampling was used 49 times up to and including Sol 1024
% Adaptive sampling was used 52 times up to and including Sol 1083
52 scans analyzed.  To our knowledge, it represents the first case of autonomous decision-making through compositional analysis performed by a spacecraft on another planet.

We first present an overview of the instrument and the onboard data analysis approach.  We then discuss the formulation and testing of rules defining targets of interest.  We then describe results and the performance of actual scans conducted on rocks on the surface of Mars. 

\subsection{The Planetary Instrument for X-ray Lithochemistry (PIXL)}
\label{spot_size}
\begin{figure}
\centering
\includegraphics[scale=0.40]{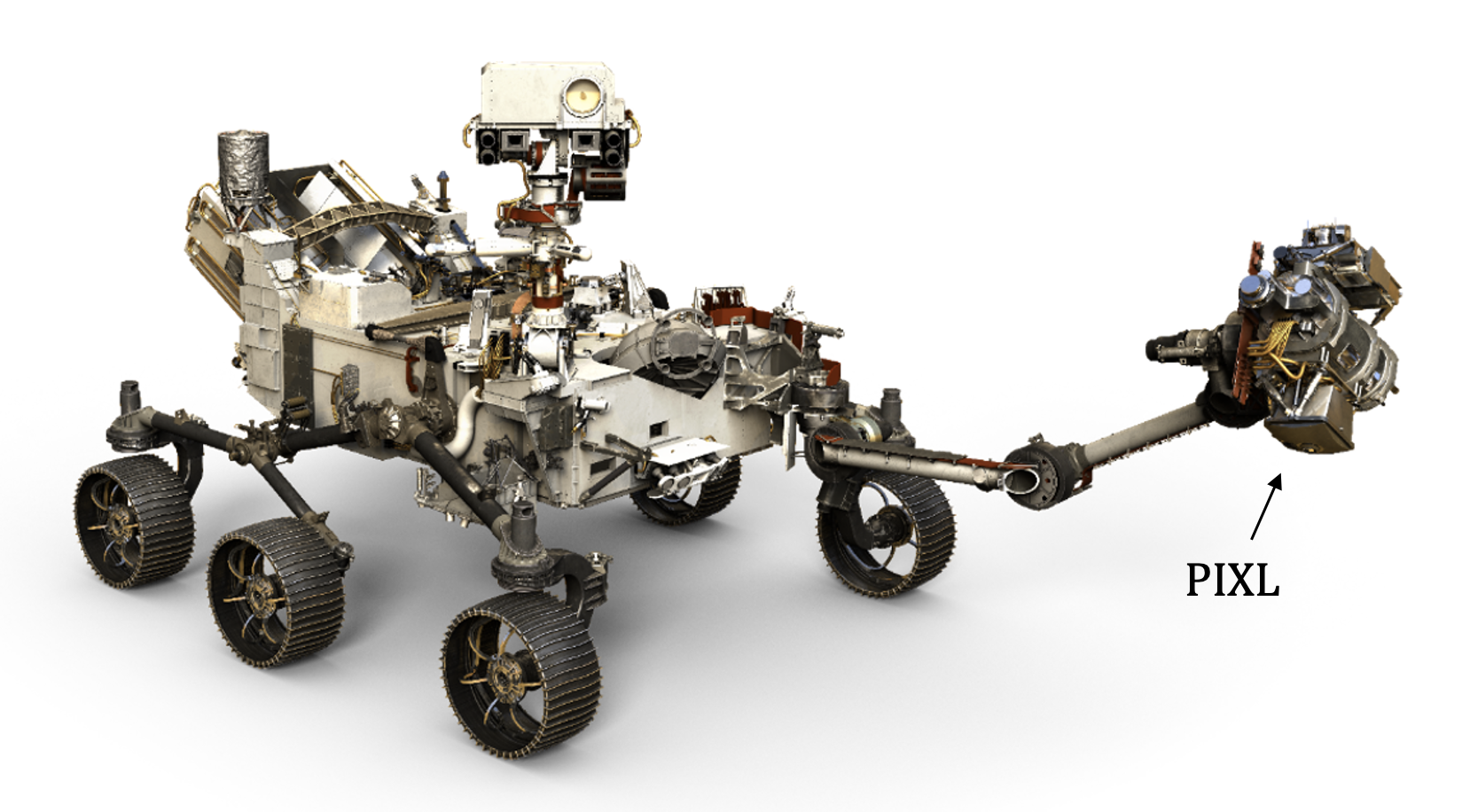}
\caption{Model of the Mars {\it Perseverance} rover showing the location of PIXL at the end of the robotic arm.}
\label{rover}
\end{figure}

PIXL is an X-ray spectrometer carried by the 
Mars {\it Perseverance} rover, mounted on the turret at the end of the rover's robotic arm, shown in Fig.\ \ref{rover}.
PIXL's primary science goal is to gather 
textural and geochemical contextual data useful for interpreting the geologic history of rocks encountered by the rover, and especially those collected for
return to Earth as part of the planned Mars Sample Return. PIXL observations have informed interpretations of the igneous emplacement history of Jezero crater floor rocks and their subsequent histories
of alteration by water \citep{liu_2022,tice_2022,farley_2022,sun_2023}.
PIXL is designed
to autonomously scan the surface of rocks in a pattern spanning up to
several centimeters in breadth and width and over a spectral energy range up to 28 keV.  This enables the detection
of many elements with atomic masses greater or equal to that of Na that are commonly found in rocks and minerals.
PIXL's nominal analytical spot size is 120 {\textmu}m full width at half maximum (FWHM) at 8 keV,
 which is roughly the same order of magnitude of the diameters of 
mineral and rock grains typically observed in sandstones and fine-grained 
igneous rocks \citep{allwood2020}.\footnote{PIXL's spot size is energy dependent: X-ray excitation and collection areas increase
for lower Z elements; i.e., Fe=140 {\textmu}m FWHM vs Na=230 {\textmu}m FWHM \citep{tice_2022}.} 

\begin{figure}
\centering
\includegraphics[scale=0.25]{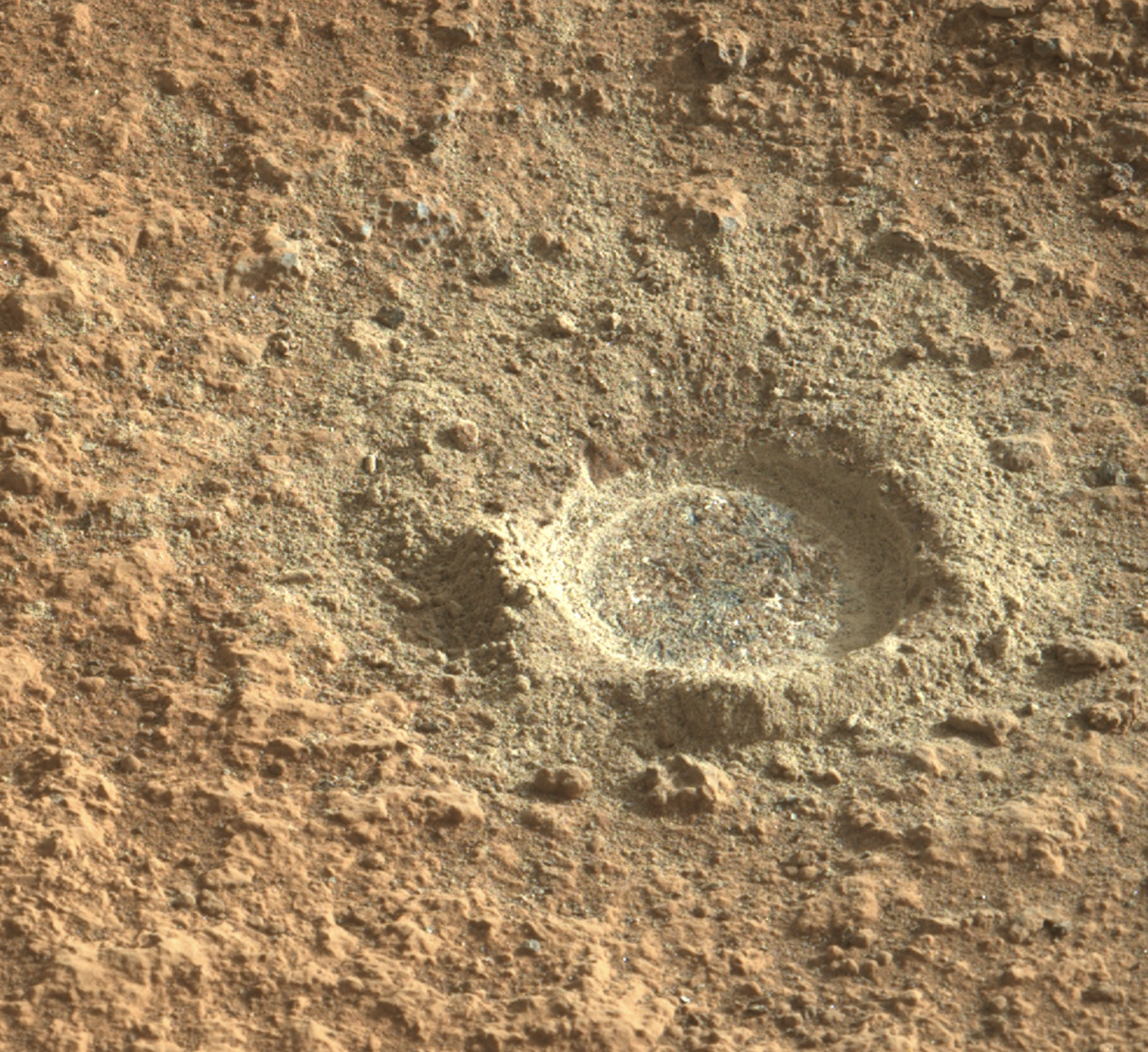}
\caption{Preparations for the PIXL scan on sol 294 (Quartier). Most PIXL scans (about 60\%)
have been conducted on an abraded patch that exposes the sub-surface of the rock 
(an additional 32\% are on natural surfaces and 8\% on regolith).
In this example an abraded patch of 50-mm diameter and approximately 7-mm deep is 
produced using the Sampling and Caching Subsystem \citep{moeller_2021}.
This patch has been cleared of dust to roughly 40-mm diameter using the Gas Dust Removal Tool.
The PIXL scan on sol 294 is 7x7 mm (placement not shown), about one seventh the diameter of the abraded patch, and has 3299 points spaced 0.125 mm
apart.
}
\label{sol294_abrasion}
\end{figure}

\begin{figure}
\centering
\includegraphics[scale=0.30]{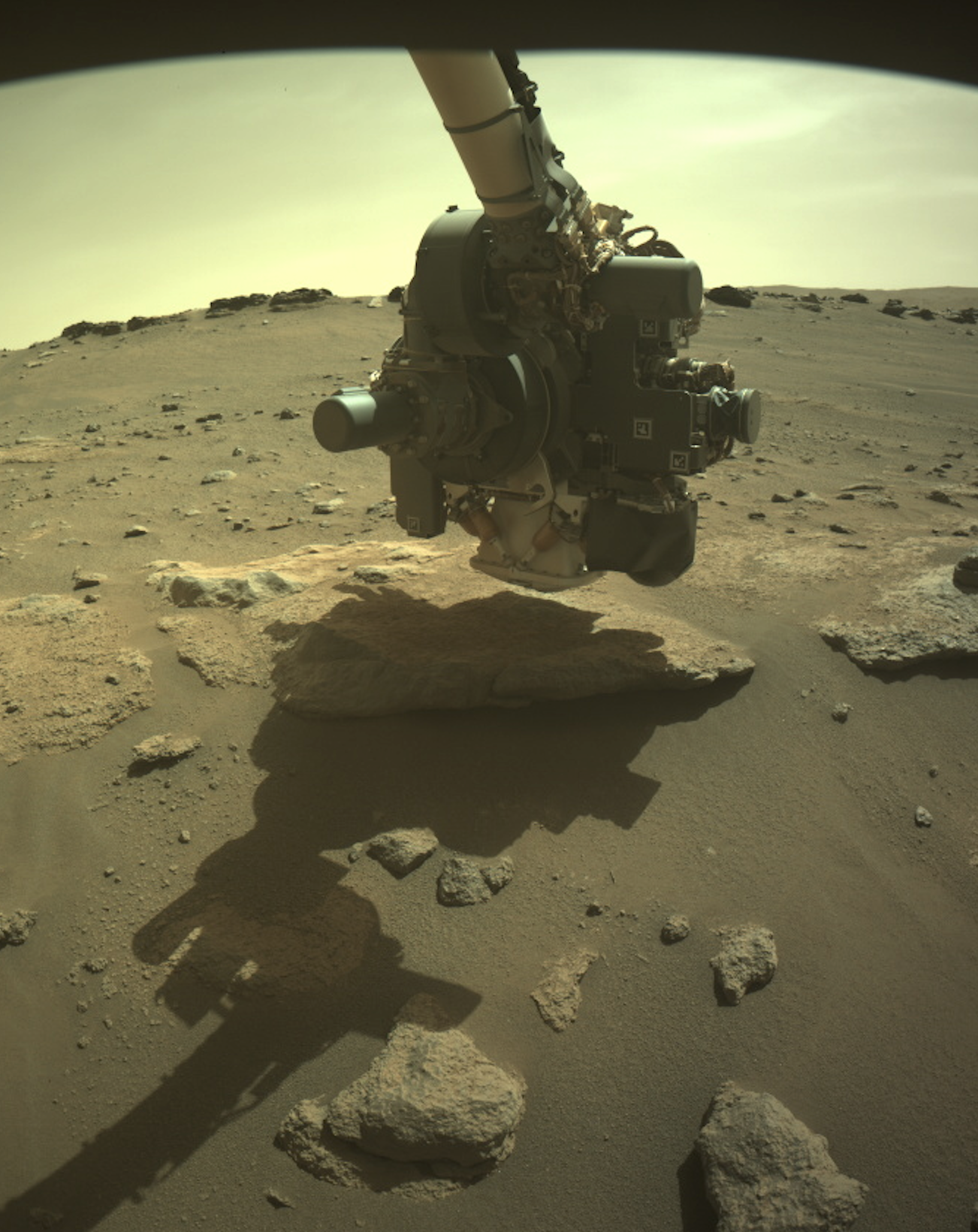}
\caption{Front Hazcam image of PIXL being placed for an overnight scan on sol 294 (Quartier).
The PIXL cover is opened and the robotic arm is used to position the instrument at a stand-off of 25.5 mm from the abraded surface.
}
\label{sol294_placement}
\end{figure}

The majority of scans are conducted as follows. 
An abrasion tool \citep{moeller_2021} is used to first expose a flat section of sub-surface rock, 
as shown in Fig.\ \ref{sol294_abrasion}. The following sol before sunset the rover arm positions PIXL over 
the target at a stand-off distance of 25.5 mm, as shown in Fig.\ \ref{sol294_placement}.
The scan is then conducted overnight.  
The scan begins and then follows a snake-like raster pattern.
At each step in the scan, the struts of PIXL's hexapod move to place PIXL at 
a new location above the surface of the target. 
An X-ray spectrum is then recorded, typically with a short 10-s dwell (integration) time. 
After acquiring the spectrum, the hexapod advances the PIXL sensor to the next physical location to repeat 
the process. 

Each PIXL raster image comprises several thousand short dwells, in total lasting 10 to 14 hours depending on the scan size, so
time is the primary limiting resource on the size of these datasets.  
Long dwells, greater than 10~s, are always desirable to accumulate more photons and improve the measurement precision, 
but it is not generally possible to program an entire raster of long dwells due to time constraints.  
Moreover, surfaces can be quite heterogeneous, with small minerals and features 
of interest distributed unevenly throughout the target matrix.  
Operators cannot anticipate where the most interesting features will be discovered
or place the instrument with sufficient accuracy to know which spectra should be lengthened.  
This motivated the team to develop PIXL's adaptive sampling capability.

\subsection{Adaptive Sampling}
Adaptive sampling aims to trigger longer dwells dynamically 
in near real-time by interpreting features in the PIXL spectra to identify compositionally valuable locations. In this 
fashion, PIXL optimizes its finite data collection time, allocating longer dwells at locations of high scientific value.

\begin{figure}
\centering
\includegraphics[scale=0.55]{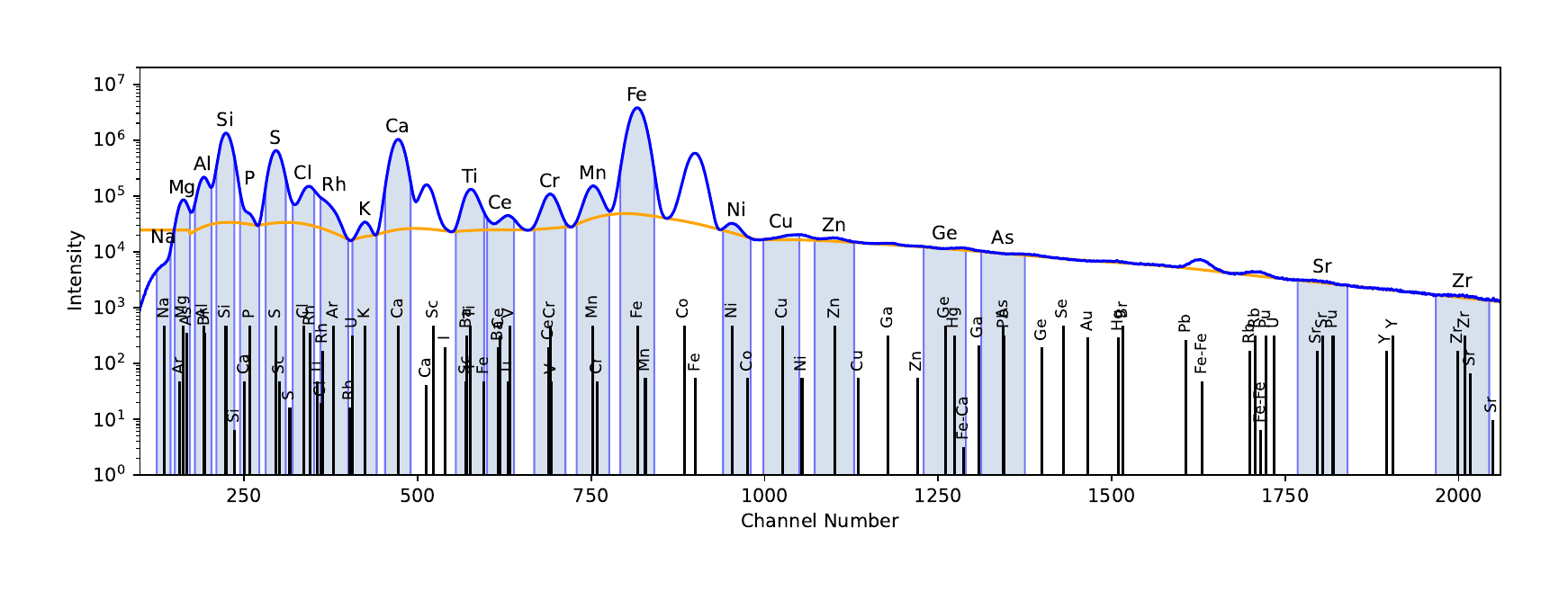}
\caption{
Pseudo-intensity channel boundaries and the locations of major peaks in the spectrum. 
The upper dark blue line is a plot of the smoothed bulk-sum spectrum of data taken on sol 865 (Dragon's Egg Rock 2). 
The estimated background is indicated by the orange line. The 22 light-blue vertical bands represent the
channels spanned by each of the 22 pseudo-intensities. For reference the locations and relative intensities of common
spectral features are indicated by the black vertical lines at the bottom of the plot.
Note that the Bremsstrahlung at channel $\sim$172 is used as a proxy for the background at lower energies, 
as discussed by \cite{Thompson2015}.
}
\label{pseudo-intensity-plot} 
\end{figure}

With adaptive sampling activated, PIXL conducts a brief analysis after acquiring each spectrum before moving to the next location.
The raw spectra from each of PIXL's two detectors are summed together, quantized in 4096 channel bins. They are then
 re-binned into 22 {\it pseudo-intensities}\, each corresponding to spectral bands spanning peaks of interest in the spectrum, as 
illustrated in Fig.\ \ref{pseudo-intensity-plot}. 
The pseudo-intensities are background-subtracted and then normalized by the background to adjust for the count-rate.
The background estimation is performed via a peak-removed non-parametric smoothing algorithm originally from
 \cite{VanGrieken2001}, adapted to PIXL by W. T. Elam \citep{Thompson2015}. 
These pseudo-intensities are proxies for the strength of the X-ray signal intensity for each element.
Figure \ref{pseudo-intensity-plot} illustrates the pseudo-intensity definitions used by PIXL, discussed further in Section \ref{pseudo-intens}.
The onboard software compares the spectrum to a library of trigger conditions, or {\it rules}, constructed in advance by the science team.
If a change or pattern of pseudo-intensities triggers an adaptive sampling rule,
then a long dwell is initiated to obtain higher signal-to-noise spectra at that location.
Once the long dwell is completed, the hexapods then move PIXL to the next step in the scan, where the process is
repeated.\footnote{There are additional parameters that can be set to control the behavior of the adaptive sampling, i.e., the total number of long dwells and
details of how the sequence of long dwells is executed; allowing for example two long dwells to occur with each trigger,
or forbidding a second trigger to occur immediately after a long dwell.}
A fixed time is allocated for adaptive sampling so that PIXL abides by schedule and power constraints imposed by the rover.
A commonly used scan is 5$\times$7 mm in area, has 2337 steps, takes about 10 hours to complete, and includes 22.5 
minutes for adaptive sampling (45 long dwells of 30 s each).

As currently implemented in PIXL, adaptive sampling doubles the signal-to-noise ratio at selected locations.
Long dwells are typically 30-s long, performed in {\it addition} 
to the standard 10-s short dwells.  The effective total integration time with 
the additional long dwell is therefore 40 s, which improves the signal-to-noise ratio 
by a factor of 2, the square root of the ratio of the respective integration times, $\sqrt{\mbox{40}/\mbox{10}}$. 
This increased sensitivity not only provides a selective increase in the sensitivity of the instrument but 
potentially improves the detection of trace elements.

\section{Approach}

\subsection{Dot Product Rules}
The adaptive sampling rules we have implemented with PIXL are almost all dot-product rules. 
These rules use a projection operator to identify pseudo-intensities that are similar to a designated template pattern. Specifically, the
rules are defined as reference vectors that are projected onto a vector of pseudo-intensities. 
If a rule vector and a pseudo-intensity vector are found to be pointing in the same direction (above a threshold value of their dot-product) 
then the rule is deemed to be satisfied and a long dwell is triggered.

The adaptive sampling flight software (FSW) is used to calculate a normalized dot-product of a rule vector 
$\vec w$ with a pseudo-intensity vector $\vec \psi$ and compare it against a threshold $\beta$. The formula used by the 
FSW is as follows:
\begin{equation}
\frac{{\vec w}\cdot{\vec \psi}}{|{\vec w}|\, |{\vec \psi}|} > \beta, \quad \mbox{where} \quad -1 \leq \beta \leq 1 \,.
\label{eqn_dotproduct}
\end{equation}
Up to 32 rules can be implemented in a single ruleset; each are evaluated sequentially by the FSW.

We have implemented two methods of creating dot-product rules for PIXL: 1) machine learning and 2) ternary compositional diagram analysis.
The method of choice depends on whether prior data exists for the targets of interest.

\subsubsection{Approach 1: Machine Learning}
For the machine learning approach we assume we have many past examples of the compositions of interest for which we 
want to trigger a long dwell.  The general approach is as follows:

\begin{enumerate}
\item  Create a training database from past examples which lists at each step in a scan (the PIXL Motor Count, or PMC) a number that 
   quantifies how interesting that point is, along with the corresponding pseudo-intensities at that step. The number, a figure of merit, 
   could be either a weight percent (wt\%) quantification of a composition or the results of a more complicated expression based on 
   those quantifications.
\item  For each composition of interest establish a threshold above which we 
would like to trigger a long dwell. 
\item  Write a linear equation as a function of the pseudo-intensities (with as yet unknown coefficients) that sums to 1 for points in the scan  
above the threshold and sums to -1 for points below the threshold. 
\item  Solve for the pseudo-intensity coefficients that best satisfy these equations. These coefficients define the rule.
\item  If the dot-product of coefficients and pseudo-intensities is greater than zero, then we say the rule is satisfied and a long dwell should be triggered.
\end{enumerate}
We can use Eq.\ \ref{eqn_dotproduct} to understand the problem as one of linear classification with a hard threshold \citep[section 19.6.4]{Russel2021}.
The rule coefficients and pseudo-intensities together provide a linear approximation of a normalized figure of merit $q$ that is 
evaluated against a normalized threshold $\beta$. 
At each PMC we have an X-ray spectrum and a set of pseudo-intensities.
If the rule vector is normalized so that $|\vec w | = 1$, we can write
the classification hypothesis $y_{\mbox{\tiny{PMC}}}$ as follows:
\begin{equation}
y_{\mbox{\tiny{PMC}}} = \begin{cases}\;\;\, 1 & \text{ if $\,q_{\mbox{\tiny{PMC}}} > \beta$} \\
                         -1 & \text{ if $\,q_{\mbox{\tiny{PMC}}} \leq \beta$} 
      \end{cases}, 
      \quad \mbox{where} \quad
      q_{\mbox{\tiny{PMC}}} = \frac{1}{|
                {\vec \psi_{\mbox{\tiny{PMC}}}}
                |} \sum_i w_i \psi_{i\, {\mbox{\tiny{PMC}}}} \, ,
\label{hypothesis}
\end{equation}
and $\psi_{i\, {\mbox{\tiny{PMC}}}}$ is the $i$th element of the 22-element pseudo-intensity vector ${\vec\psi}_{{\mbox{\tiny{PMC}}}}$
recorded at a given PMC.
The vector $\vec w$ is a set of weights (the rule) 
that defines a multi-dimensional plane that separates one class of points from all others.
If for a given ${\vec\psi}_{{\mbox{\tiny{PMC}}}}$ and $\vec w$, $y_{\mbox{\tiny{PMC}}}$ lies above the plane, 
then the classification hypothesis is 1
and a long dwell is triggered; if it falls below the plane then the hypothesis is -1 and no long dwell occurs.

In practice Eq.\ \ref{hypothesis} will yield tens of thousands of linear equations, one for every
individual PMC in the training set. 
A least-squares regression is used to solve for the weights $\vec w$ that best satisfy the
set of equations defined by $\vec y_{\mbox{\tiny{PMC}}}$ and $\vec\psi_{\mbox{\tiny{PMC}}}$.
The authors solve for these equations using a statistical gradient descent algorithm.

It should be noted that machine learning algorithms commonly pre-process the training sets to be normalized with a zero mean.
However, this pre-processing is not implemented in the PIXL FSW and the raw pseudo-intensities are used instead.

\subsubsection{Approach 2: Analysis of Ternary Compositional Diagrams}

There may however be compounds of interest for which no past observations exist, but which can nonetheless be isolated in a ternary compositional diagram.
This was our initial approach for formulating dot-product rules for PIXL on the {\it Perseverance} rover. 

A ternary compositional diagram, examples of which are shown in 
Section \ref{results}, 
 illustrates the relative abundances of three materials or compounds found in a sample. 
The diagram is triangular, and each apex is labeled with one of the three compounds. At the apex the contribution of the compound labeled there is 
100\%; on the side opposite the apex the contribution is 0\%. 
Regions of interest in a ternary diagram can be isolated as lying above or below a line representing the abundance of one compound relative to the other two. 
These abundances can be expressed as quantified weight percents, calculated by analysts on Earth using sophisticated modeling software that is not available onboard.  However, we found that it is possible to approximate the abundance calculations for specific geologic domains using linear combinations of pseudo-intensities.
The general approach is as follows:

\begin{enumerate}
\item  Create a database from past examples which lists at each step in a scan the weight percent quantification of compositions of interest along with their corresponding pseudo-intensities. 
\item  Using a least-squares fit, derive linear approximations of the weight percent quantifications as a functions of the pseudo-intensities.
\item  Draw a ternary diagram whose contributions are normalized in milli-mols that has 1) the composition of interest at the apex; 2) important distinguishing compositions in the bottom left-hand corner; and 3) a collection of other much less important distinguishing compositions in the bottom right-hand corner.
\item  Set a percent threshold for the composition of interest that isolates it from the other two groups.
\item  Express as an equation the ratio of the apex divided by the sum of the compositions in all three corners being greater than the threshold.
\item  Cross-multiply and re-arrange the equation so that it appears as an linear equation with a threshold of zero.
\item  From Step 2 above, substitute for each composition a linear equation in terms of pseudo-intensities.  
\item  Re-write the linear equation in terms of pseudo-intensities. The coefficients of the pseudo-intensities define the rule.
\end{enumerate}

We used this approach in some of the earliest dot-product rules derived for PIXL, but as the mission progressed
and many tens of thousands of PMCs of data were recorded, we re-derived all of the ternary diagram rules 
as machine-learning rules. The remainder of this paper is devoted to describing our implementation and assessment of our
  machine learning approach with PIXL.

\section{Implementation}\label{methods}

\subsection{Adaptive Sampling Simulation}

We developed a stand-alone adaptive sampling simulation environment based on PIXL FSW.\footnote{PIXL's adaptive sampling algorithm was written by 
D. R. Thompson and integrated within the FSW by M. S. Gilbert, B. J. Bornstein, and R. W. Denise.}
This included the adaptive sampling FSW and higher-level sections of FSW that implement
the decision-making logic used when PIXL steps through a scan.  We extended it to allow previously 
recorded Mars PIXL data to be re-processed and re-analyzed with new rulesets.  The re-processing
takes the raw spectra, creates pseudo-intensities based on the ruleset definitions (described below), and
then evaluates them to determine where long dwells are triggered.

We validated the simulation environment by having it re-process 
previous Mars PIXL data using the same rulesets that were on the rover, and demonstrating 
that it produced the same pseudo-intensities and the same long dwell locations as PIXL did on each scan.

We test new rulesets in simulation with previously recorded Mars data,
knowing that the behavior observed in simulation is identical to what would have been observed on Mars
had these rules been implemented. 
Rulesets that produce a large number of True Positives while yielding few False Positives
are candidates for implementation. 

\subsection{Pseudo-intensity Database}\label{pseudo-intens}
The first step in preparing a new ruleset is to create a database of pseudo-intensities.
A pseudo-intensity is a representation of part of a spectrum that records the integrated counts under a 
peak and is optionally background-subtracted and/or normalized. These pseudo-intensities are
created when data is processed in the simulation environment.
The database is created using a dummy ruleset which includes placeholder rule coefficients, 
the channel boundaries of the pseudo-intensities, and instructions 
whether background-subtraction and/or normalization is used. 

A single 4096-channel X-ray spectrum is approximated by a set of 22 pseudo-intensities.
The quantities that define a pseudo-intensity are as follows:
\begin{enumerate}
\item  Pseudo-intensity name, such as "Na", "Mg", etc.
\item  Peak location (keV), converted to a detector channel number.
\item  Peak width (keV), with the peak location converted to starting and ending channel numbers.
\item  Whether background subtraction is used or not, denoted by a 1 or 0 respectively.
\item  Whether the pseudo-intensities are normalized or not, denoted by a 1 or 0 respectively.
\end{enumerate}

Table \ref{table_pseudo_def} lists the parameters that define the pseudo-intensities at the time of 
writing.
Figure \ref{pseudo-intensity-plot} shows these pseudo-intensity bands superimposed on a bulk spectrum 
of data from sol 865.
The peak widths are proportional to the square-root of the peak channel number; in some cases the
the start and end channels were adjusted manually, based on inspection of the bulk spectra from past data.
The relationship between energy in keV and channel number for detector A is given by the following equation, where
$y$ is the energy in eV and $x$ is the channel 
number.\footnote{The
adaptive sampling algorithm uses the sum of counts from detectors A and B. Although the calibration for detector 
B is slightly different, for the purpose of 
defining the pseudo-intensity boundaries this difference is of no consequence. These 
calibrations will change slowly over the mission and must be periodically updated.} 
\begin{equation}
y = 7.86x - 18.5
\end{equation}
\begin{table}
   \centering
     \caption{Pseudo-intensity channel definitions for ruleset 230328. All of the pseudo-intensities
are background-subtracted and normalized except for Rh and Zr. The Rh signature 
originates from the PIXL X-ray tube anode,
and its pseudo-intensity is used as a measure of the background, thus the background is not subtracted in this case.
The background at Zr is very noisy and would corrupt the Zr pseudo-intensity if background-subtraction and normalization
were used, so both are disabled.}
\begin{tabular}{l c c c c c}
Element & Center & Start   & End     & Subtract   & Normalize \\
        & (keV)  & Channel & Channel & Background & \\ 
\hline
Na & 1.04 & 124  & 144 & 1 & 1 \\
Mg & 1.25 & 150  & 172 & 1 & 1 \\
Al & 1.49 & 179  & 203 & 1 & 1 \\
Si & 1.74 & 210  & 236 & 1 & 1 \\
P  & 2.01 & 244  & 272 & 1 & 1 \\
S  & 2.31 & 281  & 310 & 1 & 1 \\
Cl & 2.62 & 320  & 351 & 1 & 1 \\
Rh & 2.70 & 360  & 400 & 0 & 1 \\
K  & 3.31 & 406  & 441 & 1 & 1 \\
Ca & 3.69 & 453  & 490 & 1 & 1 \\
Ti & 4.51 & 555  & 596 & 1 & 1 \\
Ce & 4.84 & 600  & 639 & 1 & 1 \\
Cr & 5.41 & 668  & 713 & 1 & 1 \\
Mn & 5.90 & 729  & 776 & 1 & 1 \\
Fe & 6.40 & 792  & 841 & 1 & 1 \\
Ni & 7.48 & 940  & 980 & 1 & 1 \\
Cu & 8.05 & 998  & 1050 & 1 & 1 \\
Zn & 8.64 & 1072 & 1129 & 1 & 1 \\
Ge & 9.89 & 1229 & 1290 & 1 & 1 \\
As & 10.54 & 1312 & 1375 & 1 & 1 \\
Sr & 14.17 & 1768 & 1840 & 1 & 1 \\
Zr & 15.75 & 1967 & 2044 & 0 & 0 \\
  \hline
  \end{tabular}
\label{table_pseudo_def}
\end{table}
The pseudo-intensity database is created by re-processing past Mars PIXL data according to these definitions.

\subsection{Quantification Database}
The second step in preparing a new ruleset is to create a database of oxide weight percent quantifications from past scans.
Shortly after a scan has been downlinked from the rover, it becomes available to the science team in
PIXLISE \citep{nemere_2024}, a data analysis and visualization tool.
Within PIXLISE, a member of the PIXL science team runs
the PIQUANT algorithm \citep{heirwegh_2022,heirwegh_2024}
to quantify the presence of oxides found at each PMC in that scan.
PIXL has two X-ray detectors (designated as detector A and detector B), and PIQUANT is configurable to 
process data separately from each detector or with both detectors combined. 
Because the default configuration for adaptive sampling is to 
analyze the combined (A $+$ B) data, we catalog {\em combined}\ quantifications produced by PIQUANT.

\begin{figure}
\centering
\includegraphics[scale=0.40]{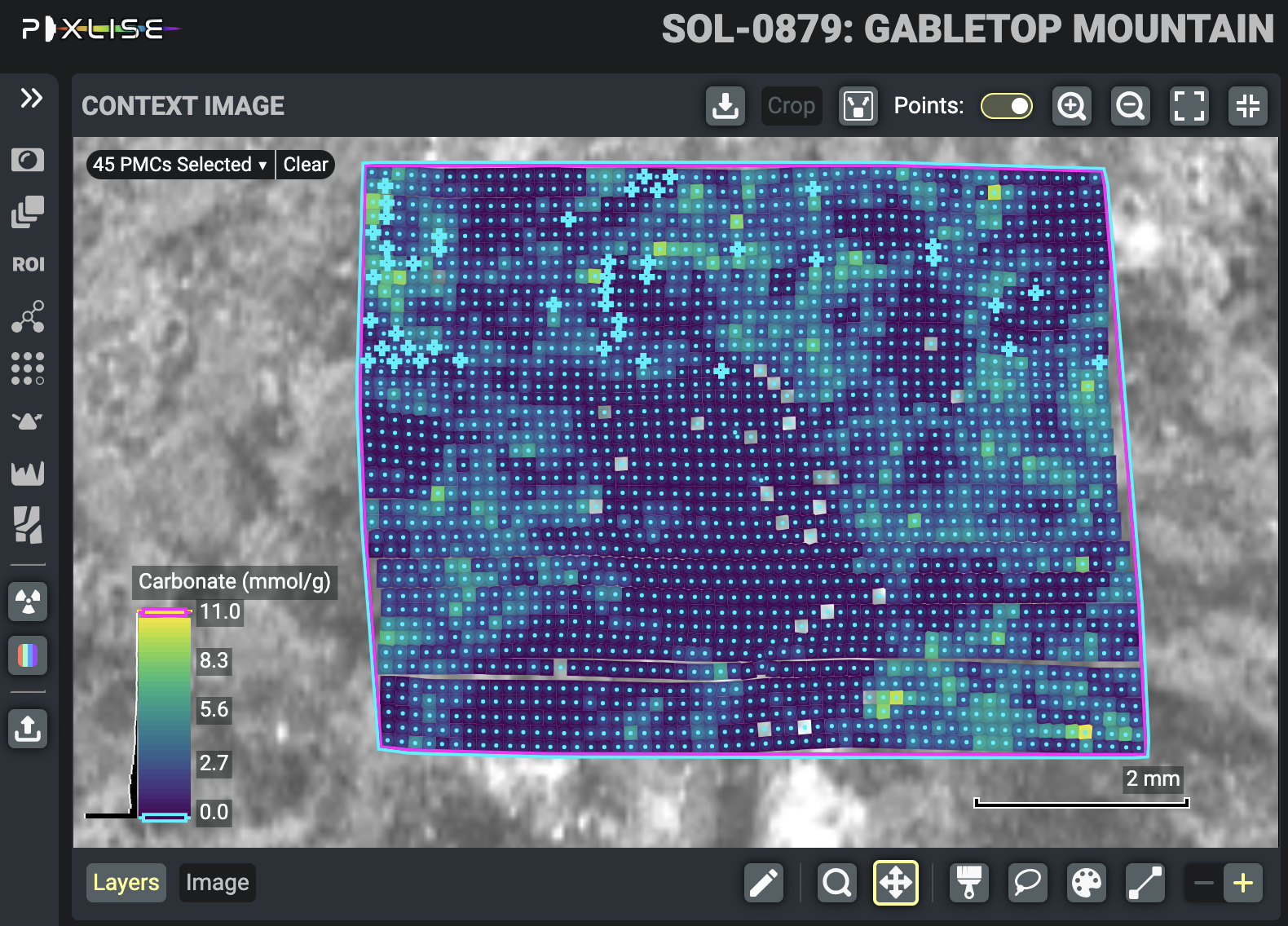}
\caption{PIXLISE view of carbonates detected on sol 879 (Gabletop Mountain).
Shown here is a compositional map of carbonates derived from PIXLISE "expressions".
The expressions are equations that can use counts from PIXL's two detectors
to estimate weight percents of compositions of interest while
taking into account the effects of diffraction. The grey-colored points
are where divide-by-zero errors have occurred and should be ignored.
}
\label{gabletop_scan}
\end{figure}

However, quantifications for carbonates are treated differently. Unlike most mineral compositions which are estimated
solely by the PIQUANT algorithm, carbonates are estimated through higher-order "expressions" in PIXLISE.
The expressions for carbonates use the PIQUANT quantifications for each separate detector but also attempt 
to remove the effects of diffraction by comparing detector counts.
Moreover the expressions for carbonates may differ from one scan to another 
depending on the composition of the substrate. We catalog the results of the agreed-upon expressions separately
from the routine PIQUANT quantifications. An example of a carbonate map is shown in Fig.\ \ref{gabletop_scan}.

\newcommand{\bpsi}{({\boldsymbol\psi})}

\subsection{Machine Learning}
The quantification database along with the pseudo-intensity database are mathematically compared to 
derive our adaptive sampling rules. 

\subsubsection{Creation of Training Datasets}

The machine learning training set for a compound of interest is created 
using thresholds of weight percent for that compound. If in past scans the weight percent at a given PMC exceeds the threshold, then 
we declare that the compound is present; conversely if it falls below the threshold, then we declare the compound is absent.

The coefficients, and therefore the performance, differ depending on the thresholds that are applied to create the training sets. Generally, the lower the threshold, the higher the True Positive Rate. However, the number of False Positives can change significantly with even a small change in threshold in the training set. Therefore to arrive at an acceptable set of coefficients, multiple
 training sets are created that each have slightly different thresholds across some nominal range. 
For example, we might create 20 training sets spanning 3.0 to 8.0 wt\% in 0.25 wt\% steps.
We then choose a training set that provides acceptable performance. 
The approach is as follows:

\begin{enumerate}
\item  Define which datasets (sols) contribute to the training: specify the most recent sol to be included and any sols to be excluded. 
\item  Establish the minimum and maximum thresholds to be tested and the number of training sets in between that will be generated.
\item For each quantity of interest and for each threshold 
create a training set.
The training set is a matrix with each row corresponding to a different PMC. 
The first column in the matrix has the value 1 or -1, depending on whether or not based on the analysis
a long dwell should be triggered at that PMC. 
The subsequent 22 columns in each row list the corresponding calculated pseudo-intensities.
\end{enumerate}

\begin{figure}
\begin{center}
\includegraphics[width=3.5in,angle=0]{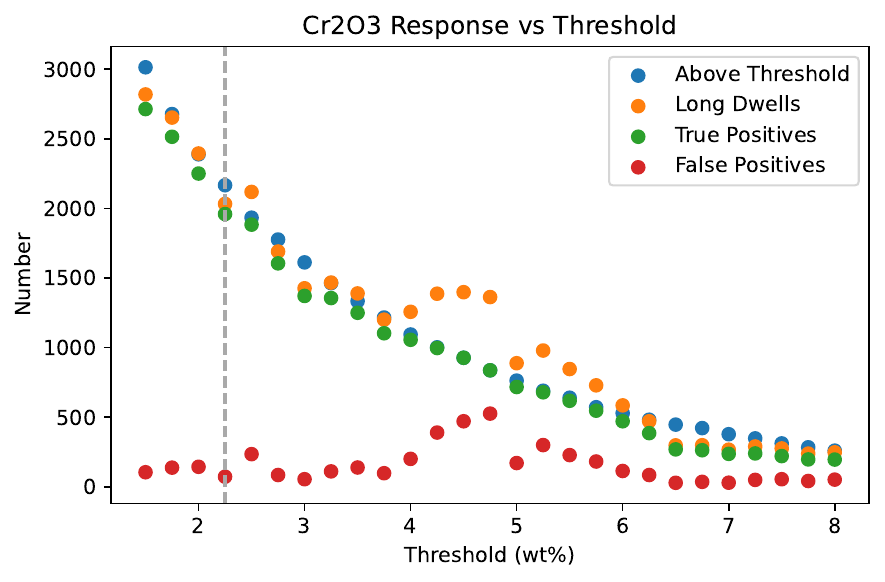}
\caption{Cr$_2$O$_3$ rule performance vs threshold. 
The performance of 27 different candidate Cr$_2$O$_3$ rules are shown, with each rule using a different
weight percent threshold. 
The number {\it Above Threshold} are the number of PMCs in the training set 
where the weight percent of Cr$_2$O$_3$ exceeds the threshold wt\%.  
The number of {\it Long Dwells} is the total number of dwells that were triggered by the rule: the 
sum of the {\it True Positives} and {\it False Positives}.
The {\it True Positives} are the number of long dwells where the weight percent
was above the threshold. The {\it False Positives} are the number of long dwells where the weight percent
was below the threshold. 
The rule coefficients for Cr$_2$O$_3$ were chosen from
the case corresponding to a threshold of 2.25 wt\%, denoted by the vertical dashed grey line, where 
the True Positives were numerous and
the False Positive were few.}
\label{cr2o3_threshold_plot}
\end{center}
\end{figure}

These files are passed along to a separate program, described below, that solves for the rule coefficients.

\subsubsection{Classification by Stochastic Gradient Descent}

Each training set is processed in the following manner to derive rule coefficients:
\begin{enumerate}
\item  Define which (if any) of the 22 pseudo-intensities should be ignored in the solution. 
\item  Read in the training data, as described previously.
\item  Delete from the matrix of equations any pseudo-intensities that should be ignored.
\item  Solve for coefficients of a linear least-squares fit to derive the best solution for the matrix of equations. 
  In our implementation we use a Stochastic Gradient Descent approach, sgdClassifier, from the scikit-learn Python library
\citep{pedregosa_2011}.
\item  Evaluate various different loss functions for Stochastic Gradient Descent (hinge, log loss, perceptron, etc.) to see which works the best. 
Coefficients of the solution with the best performance are retained, as described in the following section.
\item  Coefficients of 0.0 are inserted as part of the solution to correspond with the pseudo-intensities that were ignored.
\end{enumerate}

\subsubsection{Evaluation of Coefficients}

We evaluate the performance of a ruleset by using it in simulation to reprocess the same data it was trained on ---
all the available Mars PIXL data up until that date. The long dwells that are triggered are then compared against
the quantifications and thresholds for each compound.
As an example, Fig.\ \ref{cr2o3_threshold_plot} shows the performance of different candidate Cr$_2$O$_3$ rules, 
each created with training sets having different weight percent thresholds. 
At each PMC if the dot-product between the pseudo-intensity vector and the vector of rule coefficients 
is a number greater than zero (see Eq.\ \ref{eqn_dotproduct}) then the rule would have triggered a long dwell at that location.
The long dwells, true positives, and false positives are then evaluated. 
A ruleset is then chosen in a heuristic fashion amongst candidate rules that have a large number of 
True Positives with an acceptable number of 
False Positives.

\begin{table}
\centering
\caption{PIXL scan and adaptive sampling parameters used on Mars from sol 450 to 921.
As can be seen, adaptive sampling has been used routinely. Adjustable parameters 
include the short dwell time (Norm), the long dwell time (Dwell),
and the total number of long dwells allowed (Budget). The total number of long dwells that
were triggered (Dwells) is also listed.
}
\begin{tabular}{l l r r r r r r r}
Sol & Target Name & Samples & Surface & Ruleset & Norm & Dwell & Budget & Dwells\\
    &             &         & Type    &         & (s)  &  (s)  &        &       \\
\hline
0921 & Amherst Point 2 & 2337 & Abraded & 230328 & 10 & 30 & 45 & 45 \\
0920 & Amherst Point 1 & 2337 & Abraded & 230328 & 10 & 30 & 45 & 45 \\
0894 & Thunderbolt Peak 2 & 2581 & Abraded & 230328 & 10 & 30 & 45 & 45 \\
0887 & Thunderbolt Peak 1 & 2581 & Natural & 230328 & 10 & 30 & 60 & 60 \\
0880 & Gabletop Mnt 2 & 2581 & Abraded & 230328 & 10 & 30 & 45 & 45 \\
0879 & Gabletop Mnt 1 & 2337 & Abraded & 230328 & 10 & 30 & 45 & 45 \\
0874 & Pilot Mountain & 1377 & Natural & 230328 & 10 & 30 & 45 & 45 \\
0865 & Dragon's Egg Rock 2 & 2581 & Abraded & 230328 & 10 & 30 & 45 & 45 \\
0860 & Dragon's Egg Rock & 2337 & Abraded & 230328 & 10 & 30 & 45 & 45 \\
0852 & Lake Haiyaha 2 & 2581 & Abraded & 230328 & 10 & 30 & 45 & 45 \\
0851 & Lake Haiyaha 1 & 2337 & Abraded & 230328 & 10 & 30 & 45 & 18 \\
0847 & Lake Helene & 2337 & Natural & 230328 & 10 & 30 & 45 & 28 \\
0811 & Ouzel Falls 3 & 3333 & Abraded & 220424 & 10 & 30 & 90 & 88 \\
0790 & Ouzel Falls 2 & 825 & Abraded & 220424 & 10 & 30 & 10 & 10 \\
0789 & Ouzel Falls & 2337 & Abraded & 220424 & 10 & 30 & 45 & 25 \\
0782 & Solitude Lake & 2337 & Abraded & 220424 & 10 & 30 & 45 & 42 \\
0781 & Pipit Lake & 2337 & Natural & 220424 & 10 & 30 & 45 & 10 \\
0751 & Solva 2 & 3333 & Abraded & 220424 & 10 & 60 & 30 & 30 \\
0747 & Solva & 2337 & Abraded & 220424 & 10 & 30 & 45 & 16 \\
0711 & Cavetown & 2337 & Natural & 220424 & 10 & 30 & 45 & 11 \\
0698 & Utz Gap & 2581 & Natural & 220424 & 10 & 30 & 45 & 45 \\
0618 & Uganik Island 3 & 121 & Abraded & - & 60 & 0 & 0 & 0 \\
0617 & Uganik Island 2 & 1681 & Abraded & 221003 & 10 & 60 & 45 & 45 \\
0614 & Uganik & 3333 & Abraded & 220424 & 10 & 30 & 28 & 28 \\
0601 & Marmot Bay 2 & 1653 & Regolith & 221003 & 10 & 30 & 45 & 45 \\
0600 & Marmot Bay & 225 & Regolith & 221003 & 10 & 30 & 25 & 11 \\
0598 & Topographers Peak & 1681 & Regolith & 220424 & 10 & 30 & 45 & 17 \\
0589 & Ursus Cove 2 & 225 & Regolith & 221003 & 10 & 30 & 40 & 13 \\
0577 & Ursus Cove & 363 & Regolith & - & 10 & 0 & 0 & 0 \\
0573 & Novarupta 3 & 1681 & Abraded & 220424 & 10 & 30 & 100 & 100 \\
0570 & Novarupta 2 & 3333 & Abraded & 220424 & 10 & 30 & 100 & 100 \\
0567 & Novarupta & 1653 & Natural & 220424 & 10 & 30 & 150 & 1 \\
0560 & Chiniak Offset 2 & 1653 & Natural & 220424 & 10 & 20 & 135 & 20 \\
0558 & Chiniak Offset & 363 & Natural & 220424 & 10 & 30 & 15 & 15 \\
0507 & Berry Hollow 2 & 1653 & Abraded & 220424 & 10 & 30 & 45 & 11 \\
0505 & Berry Hollow & 3333 & Abraded & 220424 & 10 & 30 & 45 & 22 \\
0490 & Thorton Gap 2 & 363 & Abraded & 220424 & 10 & 30 & 10 & 5 \\
0484 & Thorton Gap & 3333 & Abraded & 220424 & 10 & 30 & 45 & 45 \\
0480 & Shop Hollow & 1681 & Natural & 220424 & 10 & 30 & 45 & 18 \\
0463 & Pignut Mountain & 363 & Natural & 220424 & 10 & 40 & 15 & 0 \\
0450 & Rose River Falls & 242 & Natural & 220424 & 10 & 30 & 30 & 8 \\
\hline
\end{tabular}
\label{parameters_used}
\end{table}

\begin{table}
\centering
\caption{Chronology of adaptive sampling ruleset uplinks.}
  \begin{tabular}{r r l}
Uplink Sol & Ruleset & Description \\
\hline
816 & 230328 & Cr$_2$O$_3$, TiO$_2$, P$_2$O$_5$, Carbonates, MnO, Cu, Zr  \\
% 622 & On     & Always on rule \\ 
586 & 221003 & Line scan dot-product \\
424 & 220424 & Cr$_2$O$_3$, TiO$_2$, Phosphates, Carbonate Fe Mg Mn, Carbonate Ca, Mn, Zr \\
% 300 & 211216 & Ternary dot-product  \\
% 203 & sigma  & 3$\sigma$, 4$\sigma$, 5$\sigma$, 6$\sigma$, 7$\sigma$ \\
% 13  & sigma  & 3$\sigma$ \\
  \hline
  \end{tabular}
\label{chronology}
\end{table}

\subsection{Ruleset 230328}

This paper describes the performance of ruleset 230328,\footnote{The ruleset number is a
year-month-date format (YYMMDD) denoting when the ruleset was completed: March 28, 2023.}
which was first used on the rover on sol 847 (Lake Helene).
This new ruleset was developed primarily to improve the performance of a Zr rule, which 
was designed to trigger on 7$\sigma$ changes in the Zr pseudo-intensity. 
We took the opportunity to revise the dot-product rules 
 for the detection of the other phases of interest identified by the science team 
(see Section \ref{scientific_motivation})
including carbonates, spinel minerals (identified by enrichments in Cr$_2$O$_3$ and TiO$_2$), manganese (MnO), and 
phosphate minerals (referred to as "phosphates" and identified by enrichments in P$_2$O$_5$). 
We also included a new rule to detect 6$\sigma$ changes in Cu.
A partial history of adaptive sampling use on the rover is provided in Table \ref{parameters_used}.
Additional details are provided in Table \ref{chronology}.

In the previous ruleset (220424) the Zr rule had
been triggering on noise introduced by a noisy background-subtraction and
normalization. By omitting the Zr background-subtraction and normalization 
in ruleset 230328 (see Table \ref{table_pseudo_def}), the simulated performance was greatly improved.
In the  data presented in this paper, however, there was never an opportunity for this rule to 
trigger. Nonetheless, simulations show that the Zr rule would have triggered correctly
on sol 865 (Dragon's Egg Rock 2) at PMCs 1277, 1526, and 2044, but the adaptive sampling
budget of 45 long dwells had been exhausted on that scan at PMC 1254.
The PIQUANT-derived combined weight percents of ZrO$_2$ at those PMCs are 0.74 wt\%, 1.09 wt\%, and 0.71 wt\% respectively.

These new rules were based on scan data from sols 125 through 751. 
This included scans not only on abraded surfaces, but also on 
natural surfaces and regolith. The pseudo-intensity for Zr was ignored in the creation of these dot-product rules;
the rule coefficient for Zr in each case was set to 0.0.
The resulting coefficients for the dot-product
rules are given in Table \ref{coefs_230328}. 

\begin{table}
\centering
\caption{Coefficients for ruleset 230328. Each column lists the rule coefficients 
for different dot-product rules in the ruleset. Coefficients 
are shown for rules for carbonates, Cr$_2$O$_3$, TiO$_2$, P$_2$O$_5$, and MnO.
The percentages associated with the rule denote the threshold used in the training set.
For example, 
the rule designed to detect Cr$_2$O$_3$ used a training set to detect 2.25 wt\% or greater.
The coefficients listed here are rounded to 5 decimal places for ease of inspection; the coefficients
uploaded to the rover were exactly those output from sgdClassifier, some to 9 decimal places.
The largest coefficients for each rule, those that principally determined whether
the rule will trigger or not, are shown in bold text. (Note that the Cu and Zr rules
are not dot-product rules and therefore have no coefficients listed here; they trigger from 6$\sigma$ and 7$\sigma$ 
changes in their pseudo-intensities, respectively,)
}
\begin{tabular}{l r r r r r}
Pseudo-   & Carbonates & Cr$_2$O$_3$ & TiO$_2$ & P$_2$O$_5$ & MnO\\
intensity &    4.58 wt\%  &      2.25 wt\% &  3.00 wt\% &     3.50 wt\% & 2.25 wt\% \\
\hline
Na  & 0.15851 & 0.10316 & 0.24538 & 0.04665 & -0.02502 \\
Mg  & 0.11073 & 0.00339 & -0.22796 & -0.15820 & -0.11670 \\
Al  & -0.17406 & -0.20837 & -0.07513 & 0.01868 & -0.15117 \\
Si  & -0.05041 & -0.01588 & -0.03850 & -0.09805 & -0.02721 \\
P  & -0.10058 & 0.02888 & -0.10321 & \textbf{0.95600} & -0.03069 \\
S  & -0.10094 & -0.04763 & -0.06461 & -0.02092 & -0.05203 \\
Cl  & 0.00052 & 0.03007 & -0.06779 & -0.02470 & -0.00791 \\
Rh  & -0.03586 & -0.15829 & -0.05199 & 0.00620 & -0.10985 \\
K  & 0.12458 & -0.20392 & -0.27578 & 0.14737 & -0.28563 \\
Ca  & -0.03539 & -0.05314 & -0.09512 & 0.01899 & -0.02749 \\
Ti  & -0.10685 & 0.17638 & 0.38019 & -0.02098 & -0.10216 \\
Ce  & -0.28577 & -0.13606 & \textbf{0.65148} & 0.01003 & 0.01213 \\
Cr  & -0.06706 & \textbf{0.86775} & 0.05481 & -0.02576 & -0.09002 \\
Mn  & 0.17368 & -0.17072 & -0.14616 & -0.11525 & \textbf{0.91098} \\
Fe  & 0.03227 & -0.12742 & -0.00340 & -0.00395 & -0.03076 \\
Ni  & -0.51381 & -0.06813 & -0.24518 & -0.07940 & -0.02706 \\
Cu  & -0.51252 & 0.03720 & -0.24197 & -0.05996 & -0.09403 \\
Zn  & -0.16297 & 0.06582 & 0.11837 & -0.03024 & 0.03531 \\
Ge  & \textbf{0.28414} & 0.09956 & -0.18721 & 0.01618 & 0.05607 \\
As  & -0.28140 & 0.06219 & -0.11002 & -0.01235 & -0.01691 \\
Sr  & -0.02405 & -0.00890 & -0.01712 & -0.00288 & -0.01663 \\
Zr  & 0.00000 & 0.00000 & 0.00000 & 0.00000 & 0.00000 \\
\hline
\end{tabular}
\label{coefs_230328}
\end{table}

The coefficients shown in Table \ref{coefs_230328} seem intuitively correct.
The largest coefficient of the Cr$_2$O$_3$ rule belongs to the Cr pseudo-intensity.
Similarly the largest coefficient of the P$_2$O$_5$ and MnO rules 
belong to the P and Mn pseudo-intensities respectively, as one might expect.
The largest coefficient of the TiO$_2$ rule is Ce; the primary analytical Ce line (L$\alpha$ at 4.84 keV)
overlaps the tail of the secondary Ti analytical line (K$\beta$ at 4.93 keV), thus pseudo-intensities
of Ce and Ti are intertwined and positive correlations for both with 
Ti concentration is expected (see for example Fig.\ \ref{pseudo-intensity-plot}). 
The carbonate rule is somewhat inscrutable. It does not have a single positive coefficient that stands
out from the others, but rather has positive values of a similar magnitude distributed 
across Ge, Mn, Na, K, and Mg, and negative coefficients (i.e., unlike carbonate) distributed across
Ni, Cu, Ce, As, Al, and Zn.

\section{Results: Performance Assessment}\label{results}

The performance assessment required reprocessing data on the ground for all data taken from sols 847 to 921 inclusive.
The FSW on PIXL does not report which rule in the ruleset was triggered, only that a long dwell occurred.
To understand which rule corresponded to which long dwell, the Mars data were  
re-processed using ground-based software that would run each rule
separately rather than all at once. These results were filtered to 
points within the scan where adaptive sampling was active, i.e. points within the total 
budget and points not excluded because of other sampling parameters. The number of long dwells from each
rule that were triggered between sols 847 and 894 is shown in Table \ref{disection_table}.
The PMCs associated with those long dwells are provided in the Supplemental Material.
The results are shown in Figures \ref{phosphate_plots}, \ref{cr2o3_plots}, and \ref{carbonate_plots}.
For the comparisons, long dwells caused by other rules are excluded from these plots.

\begin{table}
\centering
\caption{Number of long dwells per rule from Sol 847 to 921 with ruleset 230328. The PMCs associated with
each long dwell are provided in the Supplemental Material.}
\begin{tabular}{l l r r r r r r r r r}
Sol & Target Name       & Surface & Carbonates & Cr$_2$O$_3$ & TiO$_2$ & P$_2$O$_5$ & MnO & CuO & ZrO$_2$ & Dwells\\
\hline
0921 & Amherst Point 2     & Abraded & 10   &  34   &  0  &    0 &   0 &    1 &   0  & 45 \\
0920 & Amherst Point 1     & Abraded & 6    &  38   &  0  &    0 &   0 &    1 &   0  & 45 \\
0894 & Thunderbolt Peak 2  & Abraded & 32   &  13   &  1  &    0 &   0 &    0 &   0  & 45 \\ 
0887 & Thunderbolt Peak 1  & Natural & 40   &  17   &  2  &    0 &   0 &    1 &   0  & 60 \\
0880 & Gabletop Mnt 2      & Abraded &  5   &  33   &  0  &    2 &   1 &    5 &   0  & 45 \\
0879 & Gabletop Mnt 1      & Abraded & 29   &  13   &  2  &    1 &   0 &    2 &   0  & 45 \\
0874 & Pilot Mountain      & Natural & 31   &   6   &  5  &    0 &   0 &    3 &   0  & 45 \\
0865 & Dragon's Egg Rock 2 & Abraded &  0   &  16   &  9  &   28 &   0 &    0 &   0  & 45 \\
0860 & Dragon's Egg Rock   & Abraded &  0   &   2   &  4  &   41 &   0 &    0 &   0  & 45 \\
0852 & Lake Haiyaha 2      & Abraded & 14   &  29   &  0  &    2 &   0 &    0 &   0  & 45 \\
0851 & Lake Haiyaha 1      & Abraded &  0   &   7   &  0  &    1 &   0 &   10 &   0  & 18 \\
0847 & Lake Helene         & Natural &  0   &   7   & 21  &    0 &   0 &    0 &   0  & 28 \\
\hline
\end{tabular}
\label{disection_table}
\end{table}

Each observation typically used an adaptive sampling budget of 45 long dwells, or
approximately 22.5 minutes added to the total scan time. The plots show a 
comparison between the points that were sampled with short dwells and those
points where long dwells were triggered by the ruleset. 

The rule performance is evaluated by showing the weight percent of 
compounds at points that triggered long dwells versus the
weight percent of points that did not, subject to the filtering
described above.
The results are also shown within ternary compositional diagrams illustrating the
separation of long dwells from normally sampled points across a range
of compounds.

\subsection{Phosphate Rule}

\begin{figure}[htb]
\begin{center}
\includegraphics[width=3.5in,angle=0]{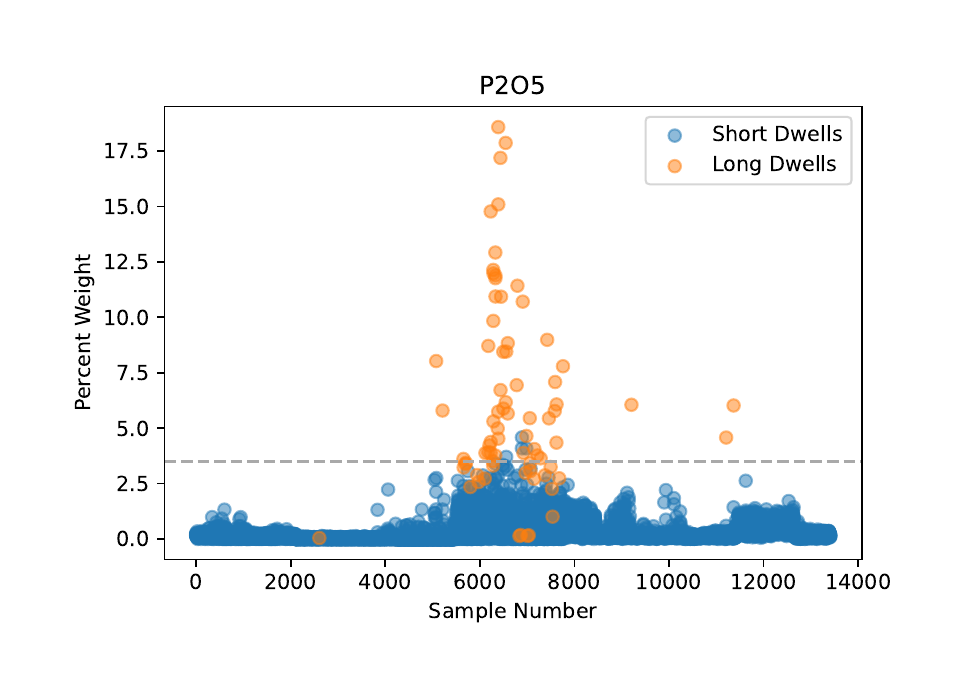}
\includegraphics[width=3.5in,angle=0]{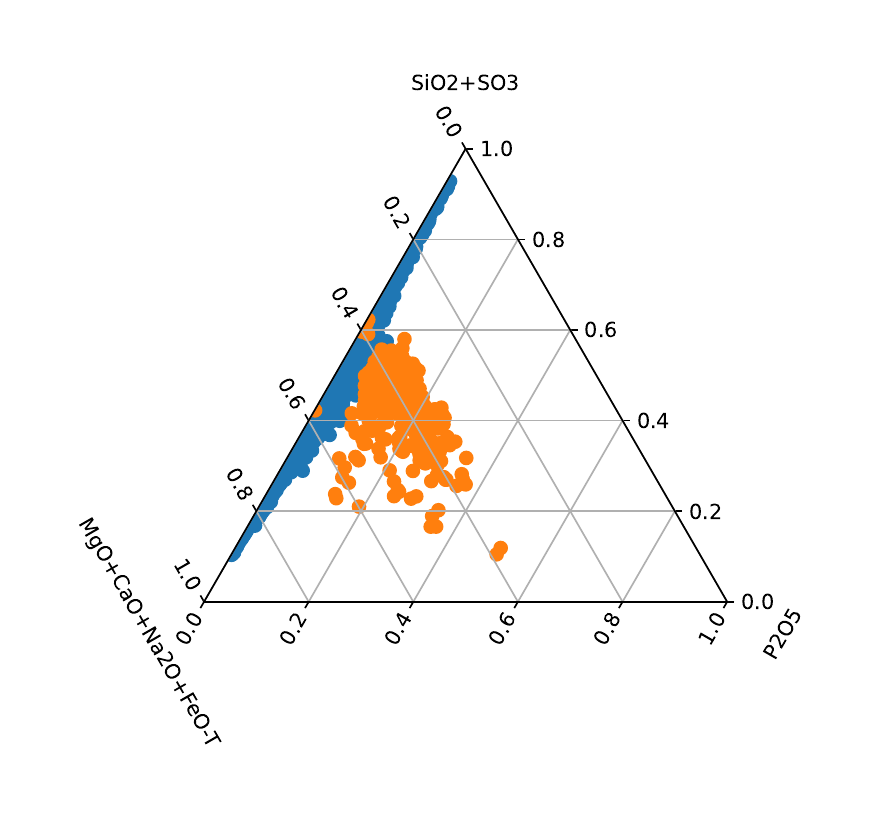}
\caption{P$_2$O$_5$ rule with a 3.5 wt\% threshold. 
The plotted data show 54 true positives, 22 false positives, and 4 false negatives, 
yielding a true positive rate of 93.1\% and a false positive rate of 0.16\% (see
Table \ref{tp_rate}).
At a small number of points there was a strong Ca escape peak within the
P pseudo-intensity band, causing the phosphate rule to trigger when there
was very little phosphorus. 
}
\label{phosphate_plots}
\end{center}
\end{figure}

The results for the phosphate rule are shown in Fig.\ \ref{phosphate_plots}.
One can see from inspection that the rule performs well. 
The separation of points in the ternary diagram is very clean.
Only a few long dwells have anomalously low weight percents of P$_2$O$_5$, 
due to the overlapping Ca escape peak and high-Ca 
within the P pseudo-intensity band at these points.

\subsection{Cr-Ti Spinel Rules}
\begin{figure}[htb]
\begin{center}
\includegraphics[width=3.5in,angle=0]{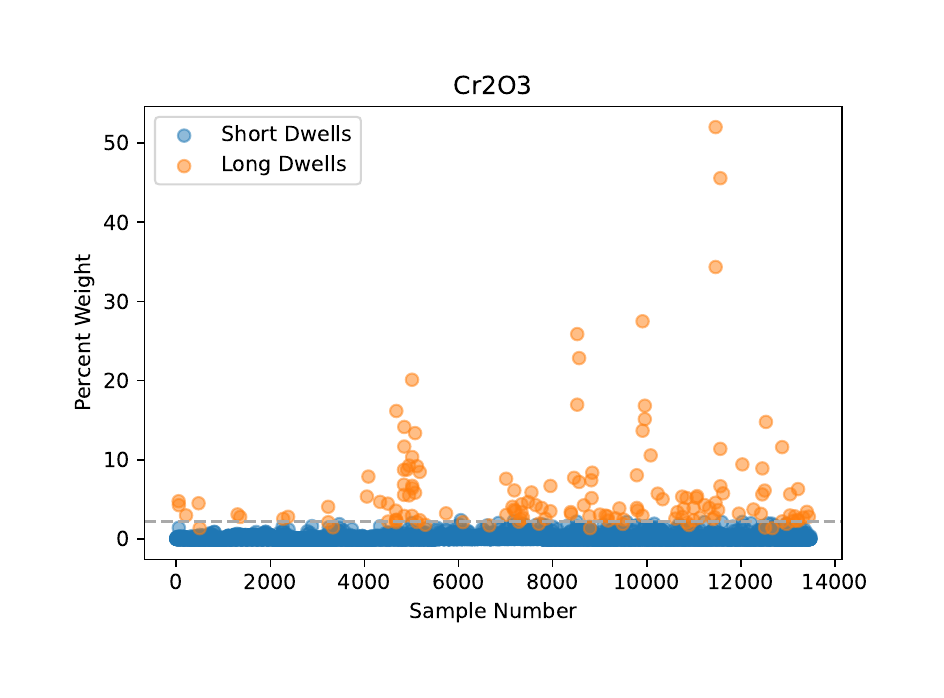}
\includegraphics[width=3.5in,angle=0]{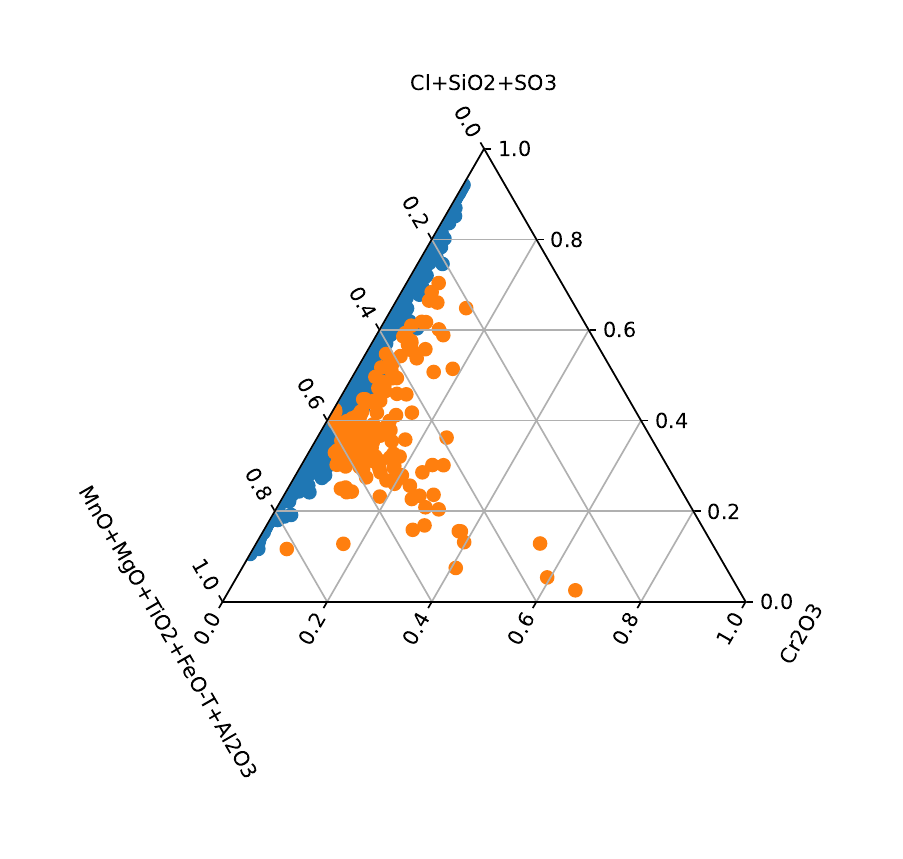}
\caption{Cr$_2$O$_3$ rule with a 2.25 wt\% threshold. 
The plotted data show 126 true positives, 19 false positives, and 2 false negatives, 
yielding a true positive rate of 98.4\% and a false positive rate of 0.14\%.}
\label{cr2o3_plots}
\end{center}
\end{figure}

Results for the Cr$_2$O$_3$ rule are shown in Fig.\ \ref{cr2o3_plots}. All of the
highest weight percent points are captured by adaptive sampling, and
only a small number of long dwells are triggered with anomalously low weight percents.
Over the range of sols for which the new ruleset was used, there was very 
little TiO$_2$ detected, and therefore these results are not shown.

\subsection{Carbonate Rule}

\begin{figure}[htb]
\begin{center}
\includegraphics[width=3.5in,angle=0]{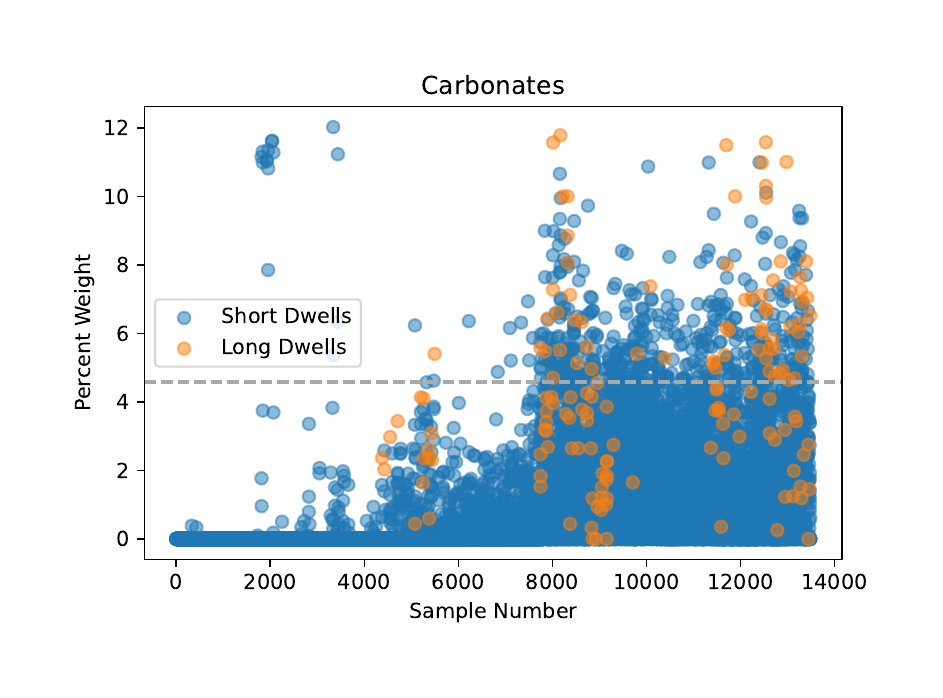}
\includegraphics[width=3.5in,angle=0]{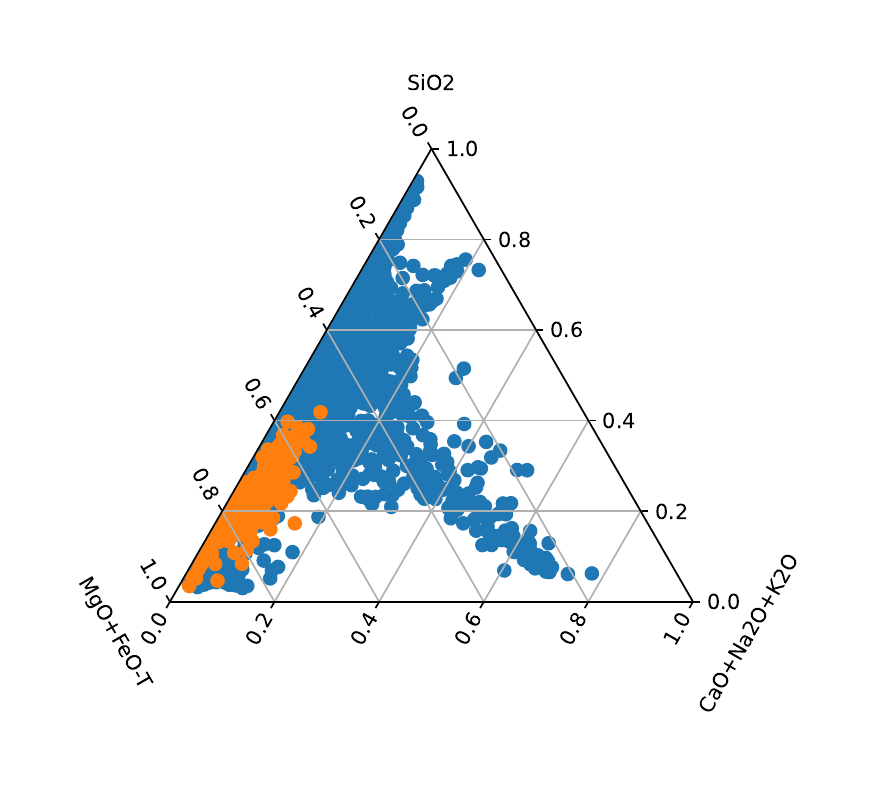}
\caption{Carbonate rule with a 4.58 wt\% threshold.
The plotted data show 68 true positives, 85 false positives, and 488 false negatives,
yielding a true positive rate of 12.2\% and a false positive rate of 0.66\%.}
\label{carbonate_plots}
\end{center}
\end{figure}

Qualitatively, through inspection of the results in PIXLISE, the carbonate rule
appears to perform well, though its true positive rate is only 12.2\%.
As can be seen in the ternary plot of Fig.\ \ref{carbonate_plots},
the rule indeed isolates carbonates as expected.
Compared to the other rules it triggers over a broader range 
of weight percent, above and below the threshold.

This discrepancy reflects the fact that PIXL does not directly observe characteristic fluorescence X-rays from C; the abundance of carbonate minerals is therefore inferred from PIXL spectra using a multi-step process using the separate spectra from detectors A and B. The adaptive sampling
algorithm, on the other hand, uses the summed data from both detectors.
The expressions on which the training was based therefore have access to information
that is hidden from the adaptive sampling algorithm. It is thus remarkable that a machine-learning based approach using a combined pseudo-intensity dataset is able to reproduce a classification based on complete spectra/quantifications from the separate detectors and expert human mineral classifications.

It is also worth noting that there are multiple types of carbonate, mixed
with various mineral compositions observed on natural and
abraded surfaces, that were included in the training set. The 
resultant rule was therefore the result of a heterogeneous mixture of data.

\subsection{Discussion: Performance Assessment}

As can be seen in Table \ref{tp_rate}, across the more than 13,300 points sampled in scans
from sol 847 to 921 inclusive, 
the True Positive Rate (TPR) is above 86\% for the Cr$_2$O$_3$, TiO$_2$, and 
P$_2$O$_5$ rules, 
and about 12\% for the carbonate rule. The MnO rule had one True Positive and one False Positive, thus a TPR of 50\%.
For every rule the False Positive Rate was below 0.7\%.

\begin{table}
\centering
\caption{Evaluation of rule classifications. 
True Positives (TP), False Negatives (FN), True Positive Rate (TPR = TP/(TP+FN)), False Positives (FP), 
True Negatives (TN), and False Positive Rate (FPR = FP/(TP+FP)) are listed below as a function of ruleset.
}
  \begin{tabular}{l c c c c c c}
Rule & TP & FN & TPR (\%) & FP & TN & FPR (\%) \\
\hline
Carbonates 4.58 wt\%  &  68 & 488 & 12.2 & 85 & 12732 & 0.66 \\
Cr$_2$O$_3$ 2.25 wt\% & 126 &   2 & 98.4 & 19 & 13332 & 0.14 \\
TiO$_2$ 3.00 wt\%  &  25 &   4 & 86.2 & 31 & 13330 & 0.23 \\
P$_2$O$_5$ 3.50 wt\%  &  54 &   4 & 93.1 & 22 & 13330 & 0.16 \\
MnO 2.25 wt\%   &   1 &   1 & 50.0 &  0 & 13333 & 0.0  \\
  \hline
  \end{tabular}
\label{tp_rate}
\end{table}

It should not be surprising that some
long dwells are triggered against samples whose weight percent falls
below the threshold.
The dot-product rules are trained using training sets whose members
have a weight percent for that compound that exceeds a specified threshold, however
the dot-product and the pseudo-intensities are all {\it normalized}, 
and so the rules trigger against a pattern of pseudo-intensities, 
not a weight percent. 

More importantly, the pseudo-intensities are each defined across
a band of X-ray energies, sometimes spanning spectral
lines of more than a single compound. 
As noted previously, the
presence of a strong Ca signature, for example, might cause the 
P$_2$O$_5$ rule to trigger. 

Diffraction from the target sample is another possible confounding factor \citep{tice_2022}.
The X-ray diffraction effects recorded in PIXL spectra are due to the 
properties of additive Bragg scattering of the PIXL source X-rays
when they encounter crystal lattice planes in mineral materials.
Diffraction can cause spurious enhancements of pseudo-intensities 
and the erroneous triggering of rules. 
Although this is a known issue, diffraction has not had a noticeable impact
on the performance of the adaptive sampling algorithm.

\section{Science Assessment}

\subsection{Scientific Motivation}
\label{scientific_motivation}

\subsubsection{Improve quantity and quality of data on smaller, rarer phases}
\label{phases_mentioned}

Longer dwells improve the quantity and quality of data on smaller, rarer phases of interest. 
The adaptive sampling rules currently on the Mars 2020 rover have been designed to 
trigger longer dwells on minor and accessory phases of interest 
(phases that comprise <10\% of a rock). 
These phases include: phosphate minerals, spinels, Zr-minerals (mainly zircon 
[ZrSiO$_4$] and/or baddeleyite [ZrO$_2$]), Mn-enrichments, and Cu-enrichments. 

Phosphate minerals, spinels, and Zr-minerals are relatively common minor 
and accessory phases in martian rocks \citep{mccubbin_2016,herd_2017,udry_2020},
but their sizes are often close to, or much less than PIXL’s spot size of $\sim$120 {\textmu}m FWHM at 8 keV 
(with larger spot sizes at lower energies, see Section \ref{spot_size}).  
Martian phosphate and spinel minerals typically range in size from 10s to 100s of {\textmu}m 
in diameter 
(i.e., \citealt{mccubbin_2016,goodrich_2003}),
while Zr-minerals are most often <10 {\textmu}m 
\citep{herd_2017}.
Due to their small volume and small sizes, even high resolution PIXL maps will 
typically only include a limited number of accessory phase analyses. In order 
to ensure good quality data are collected from these rare minerals, we target them with adaptive sampling. 

\begin{figure}
\centering
\includegraphics[scale=0.20]{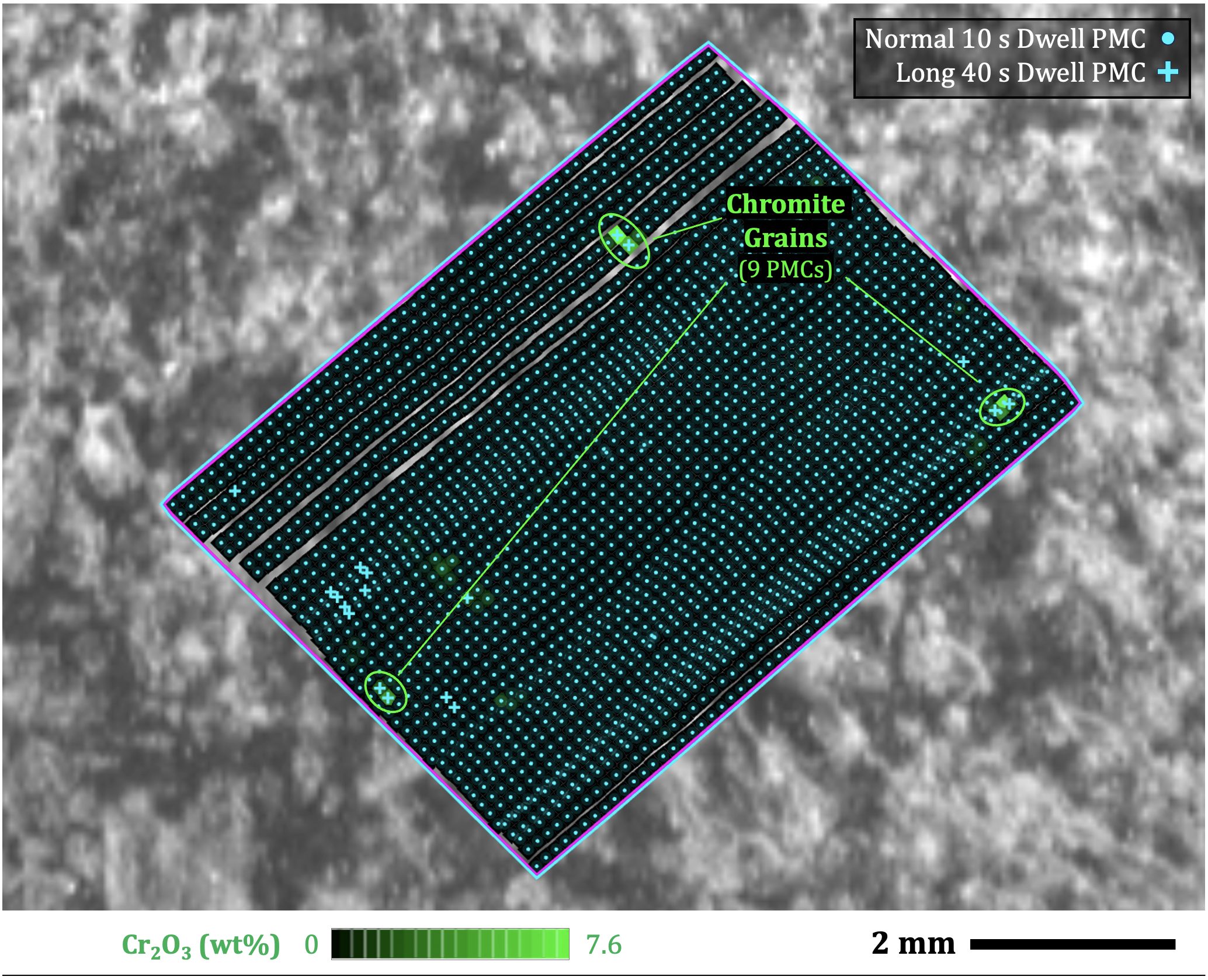}
\caption{5 x 7 mm PIXL scan of target Lake Haiyaha (sol 851) 
displaying elemental abundances of Cr$_2$O$_3$ in green. 
The scan only contains 3 Cr-rich grains (chromites) comprising 9 PMCs total 
(0.4\% of the map scan [9 PMCs/2346 PMCs]). The Chromite-bearing PMCS are defined here as PMCs with >2 wt\% Cr$_2$O$_3$.}
\label{rare_phases}
\end{figure}

Manganese is of special interest as a major constituent for several reasons.  
However, it is an element that is more typically found at a minor concentration 
level of 0.01 to 0.02 times the concentration of Fe, for which it can substitute at a low level in common minerals such as
olivine and pyroxene, including in martian meteorites \citep{papike_2009}.  
However, when Mn is found at higher 
levels than these, the minerals involved may be rare but highly 
important to the interpretation of the geochemical sample being analyzed.   
On Mars, past Mn-enrichments have been associated with sulfates 
\citep{arvidson_2016,lanza_2014},
with Cl-rich coatings \citep{berger_2019,vanbommel_2022}, 
and within phosphate minerals \citep{treiman_2023,vanbommel_2023}.
The minerals involved have been interpreted to indicate a possible higher oxidizing environment on Mars 
\citep{lanza_2015} 
or a salt precipitation sequence 
\citep{berger_2022}.
Manganese alteration minerals can provide important clues to past environmental conditions, and hence aid in the inference of environments in the past and whether they would be consistent with an enhanced or degraded level of habitability.

Copper has been found at enhanced levels in Gale Crater, Mars 
\citep{payre_2019,vanbommel_2019B,goetz_2023,forni_2024}.
When olivine weathers, it may liberate its $\sim$100 ppm Cu 
\citep{devos_2006}.
Copper is mobile under oxidizing, acidic conditions, especially at pH of 5.0 to 6.0, and has an affinity for organic matter 
\citep{devos_2006}.
Copper is an essential trace element for all known life forms.  Hence, its bio-availability has an important impact on habitability.  Furthermore, the catalytic capability of copper ions has been shown to have significant importance to a major set of chemical pathways to the prebiotic synthesis of organic molecules essential to life as we know it 
\citep{sutherland_2016,patel_2015}.
In this case, copper remains at low levels, but its associated mineral(s) can shed light on the conditions upon which it was liberated. 

\subsubsection{Improve statistics on major and trace elements}

Longer dwells should also improve the statistics on targeted phases that may contain major and trace 
elements of interest. 
When one element shows an uncommonly high level of enrichment, it is important 
to measure not only that element more accurately but also the elements which accompany it, which can aid in the
identification of the mineral involved, its origin, and geologic history.
For example, phosphate minerals are often reservoirs of a wide variety of trace 
elements such as Ce, Sr, and Y, rare earth elements, etc. of interest for petrogenetic and fluid alteration studies 
\citep{webster_2015},
but with short (i.e., normal) 10-s dwells it is unlikely that we will be able to resolve these elements due to their low 
concentrations. Ce is particularly important to quantify as its presence, even in low concentrations, can interfere with organic detections by the SHERLOC instrument (the Raman spectrometer instrument aboard the Mars 2020 rover often deployed on the same targets as PIXL 
\citep{scheller_2024}.
While a statistical analysis of Ce analyses by PIXL suggests Ce concentrations on the order of $\sim$600 ppm could be
resolvable with long dwells \citep{christian_2023}, recent work (\citealt{vanbommel_2024}, submitted) considers
calibration and spectral fitting uncertainties and strongly points to a threshold 1-2 orders of magnitude
larger, making Ce detection by PIXL unlikely.

In addition to the phases mentioned in Section \ref{phases_mentioned}, PIXL’s adaptive sampling rules have also been designed to 
trigger longer dwells on carbonate minerals in order to resolve potential trace elements. 
Trace element concentrations in carbonates can be used as markers of fluid interactions and fluid evolution.

In order to assess the effectiveness of the adaptive sampling rules thus far, we compiled all points 
identified as carbonates, spinels, and phosphates, and Mn- and Cu-enrichments in PIXL scans between sols 847 and 921 (after the 
implementation of the current adaptive sampling ruleset 230328).
We then compared how the compositions of these phases varied by producing bulk sum spectra using short dwells only, 
and a combination of short and long dwells together (see Table \ref{time_increases} for summarized results).   
Bulk sum spectra are derived through the
summation of energy-normalized spectra of similar characteristics (e.g., a region of interest with similar major element
composition), spanning not only both detectors as well as long- and short-dwells, but also multiple individual spots within a
single target scan or across multiple scans of a single target. The improved statistics offered by bulk sum spectra can
provide the statistical confidence to resolve trace elements due to a better single-to-noise ratio (e.g., \citealt{christian_2023}).

\subsection{Spinel Minerals}

Spinel minerals were identified in all scans analyzed after sol 847.
While these phases are ubiquitous, their small size and rarity resulted in relatively few 
PIXL measurements (here varying from 8 to 174 measurements/map). 
Due to the use of adaptive sampling however, the dwell time on these phases is dramatically 
increased by 39\% to 263\%, 
in comparison to integration times with normal 10~s dwells only,
allowing for better detection of major and trace elements. 

Bulk sum spectra for the spinel regions in each scan show that major element detections 
can sometimes differ when comparing normal (short) 
dwell sums versus short + long dwells sums (Table \ref{time_increases}). 
This is especially important for estimating TiO$_2$ and Cr$_2$O$_3$ concentrations, 
which are major components in these rare minerals.

\begin{figure}
\centering
\includegraphics[scale=0.45]{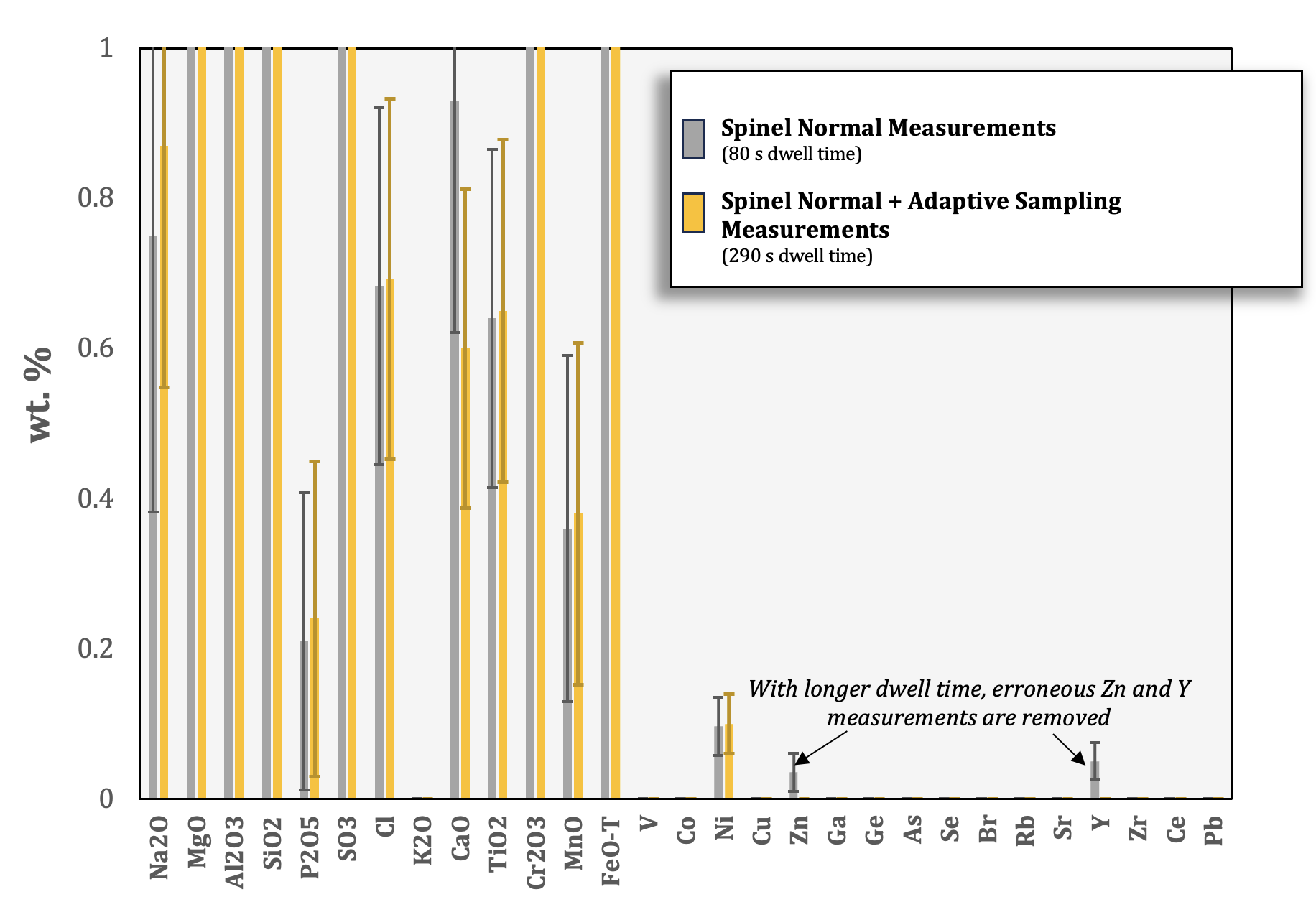}
\caption{Elemental concentrations for spinels from the Lake Helene PIXL map (sol 847) for “normal” 
dwell bulk sums, and “normal” + long dwell bulk sums. 
Erroneous quantified abundances of trace elements Zn and Y are noted to be 
below detection limit with longer dwell times.}
\label{elemental_concentrations}
\end{figure}

No trace elements of interest in the spinels were resolved from long dwells bulk sums, but 
increasing dwell time results in the removal of erroneous trace element detections in the 
higher energy part of the spectrum. 
Figure \ref{elemental_concentrations} shows 
quantified abundances of Zn and Y when considering short dwells only. An improved analysis
offered through the additional statistics provided by long dwells prompted a revised concentration
of a null result.
 In the example shown in Figure \ref{elemental_concentrations}, whereas we might have 
erroneously inferred Zn and Y enrichments in spinels in the Lake Helene (sol 847) target, data from 
the long dwells allows us to better resolve the actual concentration of these trace elements
as below detection limits.

\subsection{Phosphate Minerals}

Phosphate minerals were identified in 4 targets analyzed after sol 847. Similar to spinels, 
the  small size and rarity of the phosphate minerals resulted in relatively few PIXL 
measurements (here varying from 2 to 270 measurements/map). When long dwells triggered 
by adaptive sampling are taken into account, dwell times on phosphate minerals increases by 44\% to 300\%. 

Bulk sum spectra for the phosphate regions in each scan show that major element detections can sometimes 
differ when comparing normal dwell sums versus short + long dwells sums (Table \ref{time_increases}). 
This is especially important for estimating P$_2$O$_5$ and CaO concentrations in these rare minerals 
(the main phosphate phases on Mars are the minerals apatite (Ca$_5$(PO$_4$)$_3$(OH,F,Cl)), 
and merrillite (Ca$_9$FeMg(PO$_4$)$_7$)
(i.e., \citealt{mccubbin_2016}).

As was observed with the spinel analyses, increasing dwell times on the phosphate minerals 
resulted in the removal of erroneous quantified trace element abundances in the higher energy part of 
the spectrum (Table \ref{time_increases}). No trace elements of interest (e.g., Ce, Y, or Sr) 
were detected.

\subsection{Carbonates}

There are no changes in major element concentrations or their statistical precisions when 
comparing long and short dwells in carbonates analyzed between sols 847 and 921. 
In all of the PIXL scans considered here, carbonates are abundant across the mapped area 
so it is understandable that the estimates from long and short dwells were similar; the long 
dwells did not greatly increase the cumulative integration time for carbonates with 
increases between 2\% to 25\% (Table \ref{time_increases}).

\subsection{Mn- and Cu-enrichments}

Adaptive sampling triggered only one long dwell associated with Mn enrichments on Sol 880, 
increasing the total integration time on Mn-rich phase by 23\%. Here, bulk sum spectra for 
the Mn-rich regions shows a decrease in SiO$_2$ concentrations when comparing normal dwell sums 
versus short + long dwells sums, and no difference in trace element concentrations (Table \ref{time_increases}).

Adaptive sampling triggered long dwells associated with Cu-enrichments on Sols 851 (10 PMCs), 
874 (3 PMCs), 879 (2 PMCs), 880 (5 PMCs), and 887 (1 PMC). Careful examination of the spectra 
from these PMCs indicates that the majority of the long dwells were triggered by diffraction 
peaks (with major differences noted in the Cu peak heights between PIXL’s two detectors). 
True high-Cu concentrations triggered long dwells on 4 PMCs on Sol 851 resulting in a 300\% 
increase in dwell time. When considering short dwells only, there is a quantified 
abundance of Zr per PIQUANT modeling, however, an improved statistical analysis provided by long dwells indicates no Zr is present above detection limits.

\begin{table}
   \scriptsize
   \centering
   \begin{threeparttable}
     \caption{Summary of the dwell time increases on phases of interest targeted by adaptive sampling.}
\begin{tabular}{l l l r r r l}
    &                     &                    & Normal                & Dwell Time                      & Dwell               &  \\
    &                     & Phases of          & Dwell                 & with Long                       & Time                & Detection Changes \\
Sol & Scan                & Interest           & Time (s)              & Dwells (s)                      & Increase            & with Long Dwells\footnotemark[1] \\
\hline
921 & Amherst Point 2     & Spinels            & 1020                  & 2010                            & 97\%                & N/a \\
920 & Amherst Point 1     & Spinels            & 1740                  & 2850                            & 64\%                & N/a \\
880 & Gabletop            & Phosphates         & 20                    & 80                              & 300\%               & N/a \\
    & Mountain 2          & Spinels            & 500                   & 1430                            & 186\%               & Erroneous detection of Se removed \\
    &                     & Carbonates         & 12320                 & 12620                           & 2\%                 & N/a \\
    &                     & Mn-rich            & 130                   & 160                             & 23\%                & SiO$_2$ wt\% refined from 22.4 $\pm$1.1 to 26.8 $\pm$1.33 \\
879 & Gabletop            & Spinels            & 370                   & 760                             & 105\%               & SO$_3$ wt\% refined from 5.1 $\pm$0.3 to 3.7 $\pm$0.5 \\
    & Mountain            & Carbonates         & 8310                  & 9180                            & 10\%                & N/a \\
865 & Dragon’s Egg        & Phosphates         & 600                   & 1260                            & 110\%               & P$_2$O$_5$ wt\% refined from 10.4 $\pm$0.5 to 7.9 $\pm$0.4 \\
    & Rock 2              &                    &                       &                                 &                     & CaO wt\% refined from 10.6 $\pm$0.5 to 8.9 $\pm$0.4 \\
    &                     &                    &                       &                                 &                     & Al$_2$O$_3$ wt\% refined from 10.4 $\pm$0.5 to 11.9 $\pm$0.6 \\ 
    &                     &                    &                       &                                 &                     & Erroneous detection of Zr removed \\
    &                     & Spinels            & 500                   & 1010                            & 102\%               & SO$_3$ wt\% refined from 2.4 $\pm$0.5 to 3.5 $\pm$0.5 \\
860 & Dragon’s Egg        & Phosphates         & 2700                  & 3900                            & 44\%                & CaO wt\% refined from 17.7 $\pm$0.9 to 14.4 $\pm$0.7 \\
    & Rock                &                    &                       &                                 &                     & SO$_3$ wt\% refined from 21.9 $\pm$1.2 to 15.9 $\pm$0.8 \\
    &                     &                    &                       &                                 &                     & Al$_2$O$_3$ wt\% refined from 5.78 $\pm$0.3 to 7.87 $\pm$0.4 \\
    &                     &                    &                       &                                 &                     & MgO wt\% refined from 10.5 $\pm$0.5 to 12 $\pm$0.6 \\
    &                     &                    &                       &                                 &                     & SiO$_2$ wt\% refined from 21.8 $\pm$1.1 to 27.1 $\pm$1.4. \\
    &                     & Spinels            & 310                   & 430                             & 39\%                & SO$_3$ wt\%  refined from 14.1 $\pm$0.87 to 10.4 $\pm$0.5 \\
    &                     &                    &                       &                                 &                     & CaO wt\% refined from 10.1 $\pm$0.5 to 8.1 $\pm$0.4 \\ 
    &                     &                    &                       &                                 &                     & TiO$_2$ wt\% refined from 5.1 $\pm$0.3 to 6.3 $\pm$0.3 \\    
852 & Lake Haiyaha 2      & Phosphates         & 20                    & 80                              & 300\%               & Erroneous detection of Zr removed \\
    &                     & Spinels            & 460                   & 1390                            & 202\%               & Cr$_2$O$_3$ wt\% refined from 8.3 $\pm$0.4 to 7.3 $\pm$0.4 \\
    &                     & Carbonates         & 1310                  & 1640                            & 25\%                & N/a \\
851 & Lake Haiyaha 1      & Spinels            & 90                    & 270                             & 200\%               & Erroneous detection of Zn removed \\
    &                     & Cu-rich            & 40                    & 160                             & 300\%               & Erroneous detection of Zr removed \\
847 & Lake Helene         & Spinels            & 80                    & 290                             & 263\%               & Erroneous detections of Zn and Y removed \\
  \hline
  \end{tabular}
  \begin{tablenotes}
    \item[1] Detection changes are defined as element concentrations differing outside of error.
  \end{tablenotes}
\label{time_increases}
\end{threeparttable}
\end{table}

\subsection{Discussion: Science Assessment}

Based on our analysis of downlinked Mars data, we find that adaptive sampling succeeded in our goal of 
identifying specific high-value targets and providing better statistics on high-value compositions, 
thus improving assessments therein. The long dwells triggered by adaptive sampling sometimes 
provide better precision on major elements, but typically the values that are derived are the same, 
within error, as those obtained with short dwells alone. We can also often remove spurious detections 
of trace elements in phases of interest such as Zn, Y, Zr, and Se in spinels, phosphates, and Cu-rich 
regions that could affect scientific interpretations (Table \ref{time_increases}). On the other hand, no changes in major 
or trace element compositions were noted for carbonates when long and short dwells were compared, likely 
due to the high abundance of carbonate in PIXL scans between Sols 847 and 921. As such, future adaptive 
sampling rulesets may be revised to reduce or remove long dwells in carbonates when in carbonate-rich lithologies.   
 
The results presented in this paper suggest that detection thresholds interpreted from short dwells might 
be too low for some elements and that adaptive sampling provides more realistic limits. Adaptive sampling 
has also been shown to provide improved data on rare phases which would otherwise be difficult because 
of their small size and scarcity, with little effect on PIXL operations time and power constraints. As 
noted in the preceding sections, the typical adaptive sampling budget of 45 points on a 5 x 7 mm PIXL scan 
results in 22.5 minutes of extra scan time. If the 22.5 minutes were instead used for standard 10-s dwells, 
the area of a typical PIXL scan (2337 total analysis points) would increase by only $\sim$6\% (22.5 minutes = 135 
additional 10-s dwells). As summarized in this section and Table \ref{time_increases}, this small operational trade-off 
results in increases in dwell times on phases of interest by up to 300\%, and increases our confidence 
in the interpretation of trace element detections.

\section{Conclusion}
We have successfully demonstrated new adaptive sampling technology with PIXL on the Mars {\it Perseverance} rover.
To our knowledge, this has enabled the first autonomous decision-making based on real-time compositional analysis by an exploration spacecraft.
Almost all the rules were implemented through machine learning, trained with compositional analysis of PIXL Mars data taken earlier in the mission.
These rules are simple Linear Classifiers that use pseudo-intensities, approximations of the intensities of spectral peaks, to determine the 
presence or absence of compositions of interest. The rules were shown to work particularly well for 
spinel minerals (identified by enrichments in Cr$_2$O$_3$ and TiO$_2$) and phosphate minerals 
(identified by enrichments in P$_2$O$_5$)
 where they increased the dwell times by up to 300\% and aided in the interpretation of detections of trace elements.
Although the carbonate rule had a 12\% true positive rate and did not significantly enhance the interpretation of data, 
it is nonetheless remarkable that it worked as well as it did, 
because PIXL does not directly observe the characteristic fluorescence X-rays from C. 
The new Zr rule, despite never 
being given an opportunity to trigger, was also shown in simulations to work correctly 
and will further enhance the science return of the Mars {\it Perseverance} rover in its continuing explorations.

\section*{Acknowledgments}
The work described in this paper was partially carried out at the Jet Propulsion Laboratory, California Institute of Technology, under a contract with the National Aeronautics and Space Administration. 
We are grateful to Mars 2020 team members who participated in tactical and strategic science 
operations. We thank the support from NASA for the Phase E funding.
Particular thanks to Morgan Cable, Joel Hurowitz, and the PIXL Science Team.
We thank the following for help during laboratory development and Integration \& Test:
Mitch Au,
Hank Conley,
Marc Foote,
Christina Hernandez,
Christopher Hummel,
Joan Ervin,
Ami Kitiyakari,
Ra\'ul Romero,
Rogelio Rosas,
Michael Sondheim,
Eugenie Song,
and Michael Umana.
We are also grateful for ongoing surface operations coordinated by
Jason Van Beek,
Adrian Galvin,
James Gerhard,
and Nicholas Tallarida.
We are grateful for support of the PIXL data pipeline and the PIXLISE team; in particular
Scott Davidoff,
Peter Nemere,
and Kyle Uckert.
PRL is grateful to Felix Lawson for guidance during the early stages of software development of machine learning.

PRL, LAW, RWD, and CMH are supported by Mars 2020 Phase E funding,
DRT, MSG, and BJB were supported by Mars 2020 Phase C funding,
TVK is supported by a Canadian Space Agency M2020 Participating Scientist grant. 
SJV was supported by M2020 PS Grant \#80NSSC21K0328.

\section*{Declaration of competing interest}
The authors declare that they have no known competing financial
interests or personal relationships that could have appeared to influence
the work reported in this paper.

\section*{Data availability}
The analysis presented here is based on Localized Full Spectra (RFS), which are  
publicly available in the 
{\it PIXL Processed data collection}\, at the Geosciences Node of the
Planetary Data System (PDS; 
doi \href{https://pds.nasa.gov/ds-view/pds/viewBundle.jsp?identifier=urn\%3Anasa\%3Apds\%3Amars2020_pixl\&version=1.0}{10.17189/1522645}; 
\newline
\href{https://pds-geosciences.wustl.edu/missions/mars2020/pixl.htm}{https://pds-geosciences.wustl.edu/missions/mars2020/pixl.htm}).

\printcredits

\appendix
\section{Supplementary data}\label{sup_data}
Supplementary data to this article can be found online at TBD.

\bibliographystyle{cas-model2-names}
\bibliography{adsamp_icarus.bib}

\begin{thebibliography}{46}
\expandafter\ifx\csname natexlab\endcsname\relax\def\natexlab#1{#1}\fi
\providecommand{\url}[1]{\texttt{#1}}
\providecommand{\href}[2]{#2}
\providecommand{\path}[1]{#1}
\providecommand{\DOIprefix}{doi:}
\providecommand{\ArXivprefix}{arXiv:}
\providecommand{\URLprefix}{URL: }
\providecommand{\Pubmedprefix}{pmid:}
\providecommand{\doi}[1]{\href{http://dx.doi.org/#1}{\path{#1}}}
\providecommand{\Pubmed}[1]{\href{pmid:#1}{\path{#1}}}
\providecommand{\bibinfo}[2]{#2}
\ifx\xfnm\relax \def\xfnm[#1]{\unskip,\space#1}\fi
%Type = Article
\bibitem[{Allwood et~al.(2020)Allwood, Wade, Foote, Elam, Hurowitz, Battel,
  Dawson, Denise, Ek, Gilbert, King, Liebe, Parker, Pedersen, Randall, Sharrow,
  Sondheim, Allen, Arnett, Au, Basset, Benn, Bousman, Braun, Calvet, Clark,
  Cinquini, Conaby, Conley, Davidoff, Delaney, Denver, Diaz, Doran, Ervin,
  Evans, Flannery, Gao, Gross, Grotzinger, Hannah, Harris, Harris, He,
  Heirwegh, Hernandez, Hertzberg, Hodyss, Holden, Hummel, Jadusingh,
  Jørgensen, Kawamura, Kitiyakara, Kozaczek, Lambert, Lawson, Liu, Luchik,
  Macneal, Madsen, McLennan, McNally, Meras, Muller, Napoli, Naylor, Nemere,
  Ponomarev, Perez, Pootrakul, Romero, Rosas, Sachs, Schaefer, Schein,
  Setterfield, Singh, Song, Soria, Stek, Tallarida, Thompson, Tice, Timmermann,
  Torossian, Treiman, Tsai, Uckert, Villalvazo, Wang, Wilson, Worel, Zamani,
  Zappe, Zhong and Zimmerman}]{allwood2020}
\bibinfo{author}{Allwood, A.C.}, \bibinfo{author}{Wade, L.A.},
  \bibinfo{author}{Foote, M.C.}, \bibinfo{author}{Elam, W.T.},
  \bibinfo{author}{Hurowitz, J.A.}, \bibinfo{author}{Battel, S.},
  \bibinfo{author}{Dawson, D.E.}, \bibinfo{author}{Denise, R.W.},
  \bibinfo{author}{Ek, E.M.}, \bibinfo{author}{Gilbert, M.S.},
  \bibinfo{author}{King, M.E.}, \bibinfo{author}{Liebe, C.C.},
  \bibinfo{author}{Parker, T.}, \bibinfo{author}{Pedersen, D.A.K.},
  \bibinfo{author}{Randall, D.P.}, \bibinfo{author}{Sharrow, R.F.},
  \bibinfo{author}{Sondheim, M.E.}, \bibinfo{author}{Allen, G.},
  \bibinfo{author}{Arnett, K.}, \bibinfo{author}{Au, M.H.},
  \bibinfo{author}{Basset, C.}, \bibinfo{author}{Benn, M.},
  \bibinfo{author}{Bousman, J.C.}, \bibinfo{author}{Braun, D.},
  \bibinfo{author}{Calvet, R.J.}, \bibinfo{author}{Clark, B.},
  \bibinfo{author}{Cinquini, L.}, \bibinfo{author}{Conaby, S.},
  \bibinfo{author}{Conley, H.A.}, \bibinfo{author}{Davidoff, S.},
  \bibinfo{author}{Delaney, J.}, \bibinfo{author}{Denver, T.},
  \bibinfo{author}{Diaz, E.}, \bibinfo{author}{Doran, G.B.},
  \bibinfo{author}{Ervin, J.}, \bibinfo{author}{Evans, M.},
  \bibinfo{author}{Flannery, D.O.}, \bibinfo{author}{Gao, N.},
  \bibinfo{author}{Gross, J.}, \bibinfo{author}{Grotzinger, J.},
  \bibinfo{author}{Hannah, B.}, \bibinfo{author}{Harris, J.T.},
  \bibinfo{author}{Harris, C.M.}, \bibinfo{author}{He, Y.},
  \bibinfo{author}{Heirwegh, C.M.}, \bibinfo{author}{Hernandez, C.},
  \bibinfo{author}{Hertzberg, E.}, \bibinfo{author}{Hodyss, R.P.},
  \bibinfo{author}{Holden, J.R.}, \bibinfo{author}{Hummel, C.},
  \bibinfo{author}{Jadusingh, M.A.}, \bibinfo{author}{Jørgensen, J.L.},
  \bibinfo{author}{Kawamura, J.H.}, \bibinfo{author}{Kitiyakara, A.},
  \bibinfo{author}{Kozaczek, K.}, \bibinfo{author}{Lambert, J.L.},
  \bibinfo{author}{Lawson, P.R.}, \bibinfo{author}{Liu, Y.},
  \bibinfo{author}{Luchik, T.S.}, \bibinfo{author}{Macneal, K.M.},
  \bibinfo{author}{Madsen, S.N.}, \bibinfo{author}{McLennan, S.M.},
  \bibinfo{author}{McNally, P.}, \bibinfo{author}{Meras, P.L.},
  \bibinfo{author}{Muller, R.E.}, \bibinfo{author}{Napoli, J.},
  \bibinfo{author}{Naylor, B.J.}, \bibinfo{author}{Nemere, P.},
  \bibinfo{author}{Ponomarev, I.}, \bibinfo{author}{Perez, R.M.},
  \bibinfo{author}{Pootrakul, N.}, \bibinfo{author}{Romero, R.A.},
  \bibinfo{author}{Rosas, R.}, \bibinfo{author}{Sachs, J.},
  \bibinfo{author}{Schaefer, R.T.}, \bibinfo{author}{Schein, M.E.},
  \bibinfo{author}{Setterfield, T.P.}, \bibinfo{author}{Singh, V.},
  \bibinfo{author}{Song, E.}, \bibinfo{author}{Soria, M.M.},
  \bibinfo{author}{Stek, P.C.}, \bibinfo{author}{Tallarida, N.R.},
  \bibinfo{author}{Thompson, D.R.}, \bibinfo{author}{Tice, M.M.},
  \bibinfo{author}{Timmermann, L.}, \bibinfo{author}{Torossian, V.},
  \bibinfo{author}{Treiman, A.}, \bibinfo{author}{Tsai, S.},
  \bibinfo{author}{Uckert, K.}, \bibinfo{author}{Villalvazo, J.},
  \bibinfo{author}{Wang, M.}, \bibinfo{author}{Wilson, D.W.},
  \bibinfo{author}{Worel, S.C.}, \bibinfo{author}{Zamani, P.},
  \bibinfo{author}{Zappe, M.}, \bibinfo{author}{Zhong, F.},
  \bibinfo{author}{Zimmerman, R.}, \bibinfo{year}{2020}.
\newblock \bibinfo{title}{{PIXL: Planetary Instrument for X-Ray
  Lithochemistry}}.
\newblock \bibinfo{journal}{Space Science Reviews} \bibinfo{volume}{216},
  \bibinfo{pages}{1--132}.
\newblock \URLprefix \url{https://doi.org/10.1007/s11214-020-00767-7}.
%Type = Article
\bibitem[{Arvidson et~al.(2016)Arvidson, Squyres, Morris, Knoll, Gellert,
  Clark, Catalano, Jolliff, McLennan, Herkenhoff, VanBommel, Mittlefehldt,
  Grotzinger, Guinness, Johnson, Bell, Farrand, Stein, Fox, Golombek, Hinkle,
  Calvin and Souza}]{arvidson_2016}
\bibinfo{author}{Arvidson, R.E.}, \bibinfo{author}{Squyres, S.W.},
  \bibinfo{author}{Morris, R.V.}, \bibinfo{author}{Knoll, A.H.},
  \bibinfo{author}{Gellert, R.}, \bibinfo{author}{Clark, B.C.},
  \bibinfo{author}{Catalano, J.G.}, \bibinfo{author}{Jolliff, B.L.},
  \bibinfo{author}{McLennan, S.M.}, \bibinfo{author}{Herkenhoff, K.E.},
  \bibinfo{author}{VanBommel, S.}, \bibinfo{author}{Mittlefehldt, D.W.},
  \bibinfo{author}{Grotzinger, J.P.}, \bibinfo{author}{Guinness, E.A.},
  \bibinfo{author}{Johnson, J.R.}, \bibinfo{author}{Bell, J.F.},
  \bibinfo{author}{Farrand, W.H.}, \bibinfo{author}{Stein, N.},
  \bibinfo{author}{Fox, V.K.}, \bibinfo{author}{Golombek, M.P.},
  \bibinfo{author}{Hinkle, M.A.}, \bibinfo{author}{Calvin, W.M.},
  \bibinfo{author}{Souza, P.A.D.}, \bibinfo{year}{2016}.
\newblock \bibinfo{title}{High concentrations of manganese and sulfur in
  deposits on {Murray} ridge, {Endeavour} crater, {Mars}}.
\newblock \bibinfo{journal}{American Mineralogist} \bibinfo{volume}{101},
  \bibinfo{pages}{1389--1405}.
\newblock \URLprefix \url{https://doi.org/10.2138/am-2016-5599},
  \DOIprefix\doi{doi:10.2138/am-2016-5599}.
%Type = Inproceedings
\bibitem[{Berger et~al.(2019)Berger, King, Gellert, Clark, O’Connell-Cooper,
  Thompson, Van~Bommel and A.S.}]{berger_2019}
\bibinfo{author}{Berger, J.}, \bibinfo{author}{King, P.},
  \bibinfo{author}{Gellert, R.}, \bibinfo{author}{Clark, B.},
  \bibinfo{author}{O’Connell-Cooper, C.}, \bibinfo{author}{Thompson, L.},
  \bibinfo{author}{Van~Bommel, S.}, \bibinfo{author}{A.S., Y.},
  \bibinfo{year}{2019}.
\newblock \bibinfo{title}{{Manganese enrichment pathways relevant to {Gale}
  crater, {Mars}: Evaporative concentration and chlorine-induced
  precipitation}}, in: \bibinfo{booktitle}{{Lunar and Planetary Science
  Conference}}, \bibinfo{address}{The Woodlands (Texas), United States}.
\newblock \URLprefix \url{https://hal.science/hal-04446048}.
%Type = Article
\bibitem[{Berger et~al.(2022)Berger, King, Gellert, Clark, Flood, McCraig,
  Ming, O’Connell-Cooper, Schmidt, Thompson, VanBommel, Wilhelm and
  Yen}]{berger_2022}
\bibinfo{author}{Berger, J.A.}, \bibinfo{author}{King, P.L.},
  \bibinfo{author}{Gellert, R.}, \bibinfo{author}{Clark, B.C.},
  \bibinfo{author}{Flood, V.A.}, \bibinfo{author}{McCraig, M.A.},
  \bibinfo{author}{Ming, D.W.}, \bibinfo{author}{O’Connell-Cooper, C.D.},
  \bibinfo{author}{Schmidt, M.E.}, \bibinfo{author}{Thompson, L.M.},
  \bibinfo{author}{VanBommel, S.J.V.}, \bibinfo{author}{Wilhelm, B.},
  \bibinfo{author}{Yen, A.S.}, \bibinfo{year}{2022}.
\newblock \bibinfo{title}{Manganese mobility in {Gale} crater, {Mars}: Leached
  bedrock and localized enrichments}.
\newblock \bibinfo{journal}{Journal of Geophysical Research: Planets}
  \bibinfo{volume}{127}, \bibinfo{pages}{e2021JE007171}.
\newblock \URLprefix
  \url{https://agupubs.onlinelibrary.wiley.com/doi/abs/10.1029/2021JE007171},
  \DOIprefix\doi{https://doi.org/10.1029/2021JE007171},
  \href{http://arxiv.org/abs/https://agupubs.onlinelibrary.wiley.com/doi/pdf/10.1029/2021JE007171}{\tt
  arXiv:https://agupubs.onlinelibrary.wiley.com/doi/pdf/10.1029/2021JE007171}.
  \bibinfo{note}{e2021JE007171 2021JE007171}.
%Type = Inproceedings
\bibitem[{Candela et~al.(2017)Candela, Thompson, Dobrea and
  Wettergreen}]{candela2017}
\bibinfo{author}{Candela, A.}, \bibinfo{author}{Thompson, D.},
  \bibinfo{author}{Dobrea, E.N.}, \bibinfo{author}{Wettergreen, D.},
  \bibinfo{year}{2017}.
\newblock \bibinfo{title}{Planetary robotic exploration driven by science
  hypotheses for geologic mapping}, in: \bibinfo{booktitle}{2017 IEEE/RSJ
  International Conference on Intelligent Robots and Systems (IROS)},
  \bibinfo{organization}{IEEE}. pp. \bibinfo{pages}{3811--3818}.
\newblock \DOIprefix\doi{10.1109/IROS.2017.8206231}.
%Type = Article
\bibitem[{Castano et~al.(2008)Castano, Fukunaga, Biesiadecki, Neakrase,
  Whelley, Greeley, Lemmon, Castano and Chien}]{castano2008}
\bibinfo{author}{Castano, A.}, \bibinfo{author}{Fukunaga, A.},
  \bibinfo{author}{Biesiadecki, J.}, \bibinfo{author}{Neakrase, L.},
  \bibinfo{author}{Whelley, P.}, \bibinfo{author}{Greeley, R.},
  \bibinfo{author}{Lemmon, M.}, \bibinfo{author}{Castano, R.},
  \bibinfo{author}{Chien, S.}, \bibinfo{year}{2008}.
\newblock \bibinfo{title}{Automatic detection of dust devils and clouds on
  {Mars}}.
\newblock \bibinfo{journal}{Machine Vision and Applications}
  \bibinfo{volume}{19}, \bibinfo{pages}{467--482}.
\newblock \URLprefix \url{https://doi.org/10.1007/s00138-007-0081-3}.
%Type = Article
\bibitem[{Christian et~al.(2023)Christian, VanBommel, Elam, Ganly, Hurowitz,
  Heirwegh, Allwood, Clark, Kizovski and Knight}]{christian_2023}
\bibinfo{author}{Christian, J.R.}, \bibinfo{author}{VanBommel, S.J.},
  \bibinfo{author}{Elam, W.T.}, \bibinfo{author}{Ganly, B.},
  \bibinfo{author}{Hurowitz, J.A.}, \bibinfo{author}{Heirwegh, C.M.},
  \bibinfo{author}{Allwood, A.C.}, \bibinfo{author}{Clark, B.C.},
  \bibinfo{author}{Kizovski, T.V.}, \bibinfo{author}{Knight, A.L.},
  \bibinfo{year}{2023}.
\newblock \bibinfo{title}{Statistical characterization of pixl trace element
  detection limits}.
\newblock \bibinfo{journal}{Acta Astronautica} \bibinfo{volume}{212},
  \bibinfo{pages}{534--540}.
\newblock \URLprefix
  \url{https://www.sciencedirect.com/science/article/pii/S0094576523004393},
  \DOIprefix\doi{https://doi.org/10.1016/j.actaastro.2023.08.032}.
%Type = Inbook
\bibitem[{De~Vos et~al.(2006)De~Vos, Gregorauskiene, Marsina, Salminen,
  Salpeteur, Tarvainen, O’Connor, Demetriades, Pirc, Batista, Bidovec,
  Bel-lan, Birke, Breward, De~Vivo, Duris, Halamic, Klein, Lima, Locutura, Lis,
  Mazreku, Ottesen, Pasieczna, Petersell, Reeder, Siewers and
  Slaninka}]{devos_2006}
\bibinfo{author}{De~Vos, W.}, \bibinfo{author}{Gregorauskiene, V.},
  \bibinfo{author}{Marsina, K.}, \bibinfo{author}{Salminen, R.},
  \bibinfo{author}{Salpeteur, I.}, \bibinfo{author}{Tarvainen, T.},
  \bibinfo{author}{O’Connor, P.}, \bibinfo{author}{Demetriades, A.},
  \bibinfo{author}{Pirc, S.}, \bibinfo{author}{Batista, M.},
  \bibinfo{author}{Bidovec, M.}, \bibinfo{author}{Bel-lan, A.},
  \bibinfo{author}{Birke, M.}, \bibinfo{author}{Breward, N.},
  \bibinfo{author}{De~Vivo, B.}, \bibinfo{author}{Duris, M.},
  \bibinfo{author}{Halamic, J.}, \bibinfo{author}{Klein, P.},
  \bibinfo{author}{Lima, A.}, \bibinfo{author}{Locutura, J.},
  \bibinfo{author}{Lis, J.}, \bibinfo{author}{Mazreku, A.},
  \bibinfo{author}{Ottesen, R.}, \bibinfo{author}{Pasieczna, A.},
  \bibinfo{author}{Petersell, V.}, \bibinfo{author}{Reeder, S.},
  \bibinfo{author}{Siewers, U.}, \bibinfo{author}{Slaninka, I.},
  \bibinfo{year}{2006}.
\newblock \bibinfo{title}{Distribution of elements in subsoil and topsoil}.
\newblock pp. \bibinfo{pages}{21--29}.
\newblock \URLprefix
  \url{http://weppi.gtk.fi/publ/foregsatlas/article2.php?id=21}.
%Type = Article
\bibitem[{Estlin et~al.(2012)Estlin, Bornstein, Gaines, Anderson, Thompson,
  Burl, Castano and Judd}]{estlin2012}
\bibinfo{author}{Estlin, T.A.}, \bibinfo{author}{Bornstein, B.J.},
  \bibinfo{author}{Gaines, D.M.}, \bibinfo{author}{Anderson, R.C.},
  \bibinfo{author}{Thompson, D.R.}, \bibinfo{author}{Burl, M.},
  \bibinfo{author}{Castano, R.}, \bibinfo{author}{Judd, M.},
  \bibinfo{year}{2012}.
\newblock \bibinfo{title}{{AEGIS} automated science targeting for the {MER}
  {Opportunity} rover}.
\newblock \bibinfo{journal}{ACM Transactions on Intelligent Systems and
  Technology (TIST)} \bibinfo{volume}{3}, \bibinfo{pages}{1--19}.
\newblock \URLprefix \url{https://doi.org/10.1145/2168752.2168764},
  \DOIprefix\doi{10.1145/2168752.2168764}.
%Type = Article
\bibitem[{Farley et~al.(2022)Farley, Stack, Shuster, Horgan, Hurowitz, Tarnas,
  Simon, Sun, Scheller, Moore, McLennan, Vasconcelos, Wiens, Treiman, Mayhew,
  Beyssac, Kizovski, Tosca, Williford, Crumpler, Beegle, Bell, Ehlmann, Liu,
  Maki, Schmidt, Allwood, Amundsen, Bhartia, Bosak, Brown, Clark, Cousin,
  Forni, Gabriel, Goreva, Gupta, Hamran, Herd, Hickman-Lewis, Johnson, Kah,
  Kelemen, Kinch, Mandon, Mangold, Quantin-Nataf, Rice, Russell, Sharma,
  Siljeström, Steele, Sullivan, Wadhwa, Weiss, Williams, Wogsland, Willis,
  Acosta-Maeda, Beck, Benzerara, Bernard, Burton, Cardarelli, Chide, Clavé,
  Cloutis, Cohen, Czaja, Debaille, Dehouck, Fairén, Flannery, Fleron, Fouchet,
  Frydenvang, Garczynski, Gibbons, Hausrath, Hayes, Henneke, Jørgensen, Kelly,
  Lasue, Mouélic, Madariaga, Maurice, Merusi, Meslin, Milkovich, Million,
  Moeller, Núñez, Ollila, Paar, Paige, Pedersen, Pilleri, Pilorget, Pinet,
  Rice, Royer, Sautter, Schulte, Sephton, Sharma, Sholes, Spanovich, Clair,
  Tate, Uckert, VanBommel, Yanchilina and Zorzano}]{farley_2022}
\bibinfo{author}{Farley, K.A.}, \bibinfo{author}{Stack, K.M.},
  \bibinfo{author}{Shuster, D.L.}, \bibinfo{author}{Horgan, B.H.N.},
  \bibinfo{author}{Hurowitz, J.A.}, \bibinfo{author}{Tarnas, J.D.},
  \bibinfo{author}{Simon, J.I.}, \bibinfo{author}{Sun, V.Z.},
  \bibinfo{author}{Scheller, E.L.}, \bibinfo{author}{Moore, K.R.},
  \bibinfo{author}{McLennan, S.M.}, \bibinfo{author}{Vasconcelos, P.M.},
  \bibinfo{author}{Wiens, R.C.}, \bibinfo{author}{Treiman, A.H.},
  \bibinfo{author}{Mayhew, L.E.}, \bibinfo{author}{Beyssac, O.},
  \bibinfo{author}{Kizovski, T.V.}, \bibinfo{author}{Tosca, N.J.},
  \bibinfo{author}{Williford, K.H.}, \bibinfo{author}{Crumpler, L.S.},
  \bibinfo{author}{Beegle, L.W.}, \bibinfo{author}{Bell, J.F.},
  \bibinfo{author}{Ehlmann, B.L.}, \bibinfo{author}{Liu, Y.},
  \bibinfo{author}{Maki, J.N.}, \bibinfo{author}{Schmidt, M.E.},
  \bibinfo{author}{Allwood, A.C.}, \bibinfo{author}{Amundsen, H.E.F.},
  \bibinfo{author}{Bhartia, R.}, \bibinfo{author}{Bosak, T.},
  \bibinfo{author}{Brown, A.J.}, \bibinfo{author}{Clark, B.C.},
  \bibinfo{author}{Cousin, A.}, \bibinfo{author}{Forni, O.},
  \bibinfo{author}{Gabriel, T.S.J.}, \bibinfo{author}{Goreva, Y.},
  \bibinfo{author}{Gupta, S.}, \bibinfo{author}{Hamran, S.E.},
  \bibinfo{author}{Herd, C.D.K.}, \bibinfo{author}{Hickman-Lewis, K.},
  \bibinfo{author}{Johnson, J.R.}, \bibinfo{author}{Kah, L.C.},
  \bibinfo{author}{Kelemen, P.B.}, \bibinfo{author}{Kinch, K.B.},
  \bibinfo{author}{Mandon, L.}, \bibinfo{author}{Mangold, N.},
  \bibinfo{author}{Quantin-Nataf, C.}, \bibinfo{author}{Rice, M.S.},
  \bibinfo{author}{Russell, P.S.}, \bibinfo{author}{Sharma, S.},
  \bibinfo{author}{Siljeström, S.}, \bibinfo{author}{Steele, A.},
  \bibinfo{author}{Sullivan, R.}, \bibinfo{author}{Wadhwa, M.},
  \bibinfo{author}{Weiss, B.P.}, \bibinfo{author}{Williams, A.J.},
  \bibinfo{author}{Wogsland, B.V.}, \bibinfo{author}{Willis, P.A.},
  \bibinfo{author}{Acosta-Maeda, T.A.}, \bibinfo{author}{Beck, P.},
  \bibinfo{author}{Benzerara, K.}, \bibinfo{author}{Bernard, S.},
  \bibinfo{author}{Burton, A.S.}, \bibinfo{author}{Cardarelli, E.L.},
  \bibinfo{author}{Chide, B.}, \bibinfo{author}{Clavé, E.},
  \bibinfo{author}{Cloutis, E.A.}, \bibinfo{author}{Cohen, B.A.},
  \bibinfo{author}{Czaja, A.D.}, \bibinfo{author}{Debaille, V.},
  \bibinfo{author}{Dehouck, E.}, \bibinfo{author}{Fairén, A.G.},
  \bibinfo{author}{Flannery, D.T.}, \bibinfo{author}{Fleron, S.Z.},
  \bibinfo{author}{Fouchet, T.}, \bibinfo{author}{Frydenvang, J.},
  \bibinfo{author}{Garczynski, B.J.}, \bibinfo{author}{Gibbons, E.F.},
  \bibinfo{author}{Hausrath, E.M.}, \bibinfo{author}{Hayes, A.G.},
  \bibinfo{author}{Henneke, J.}, \bibinfo{author}{Jørgensen, J.L.},
  \bibinfo{author}{Kelly, E.M.}, \bibinfo{author}{Lasue, J.},
  \bibinfo{author}{Mouélic, S.L.}, \bibinfo{author}{Madariaga, J.M.},
  \bibinfo{author}{Maurice, S.}, \bibinfo{author}{Merusi, M.},
  \bibinfo{author}{Meslin, P.Y.}, \bibinfo{author}{Milkovich, S.M.},
  \bibinfo{author}{Million, C.C.}, \bibinfo{author}{Moeller, R.C.},
  \bibinfo{author}{Núñez, J.I.}, \bibinfo{author}{Ollila, A.M.},
  \bibinfo{author}{Paar, G.}, \bibinfo{author}{Paige, D.A.},
  \bibinfo{author}{Pedersen, D.A.K.}, \bibinfo{author}{Pilleri, P.},
  \bibinfo{author}{Pilorget, C.}, \bibinfo{author}{Pinet, P.C.},
  \bibinfo{author}{Rice, J.W.}, \bibinfo{author}{Royer, C.},
  \bibinfo{author}{Sautter, V.}, \bibinfo{author}{Schulte, M.},
  \bibinfo{author}{Sephton, M.A.}, \bibinfo{author}{Sharma, S.K.},
  \bibinfo{author}{Sholes, S.F.}, \bibinfo{author}{Spanovich, N.},
  \bibinfo{author}{Clair, M.S.}, \bibinfo{author}{Tate, C.D.},
  \bibinfo{author}{Uckert, K.}, \bibinfo{author}{VanBommel, S.J.},
  \bibinfo{author}{Yanchilina, A.G.}, \bibinfo{author}{Zorzano, M.P.},
  \bibinfo{year}{2022}.
\newblock \bibinfo{title}{Aqueously altered igneous rocks sampled on the floor
  of jezero crater, mars}.
\newblock \bibinfo{journal}{Science} \bibinfo{volume}{377},
  \bibinfo{pages}{eabo2196}.
\newblock \URLprefix
  \url{https://www.science.org/doi/abs/10.1126/science.abo2196},
  \DOIprefix\doi{10.1126/science.abo2196},
  \href{http://arxiv.org/abs/https://www.science.org/doi/pdf/10.1126/science.abo2196}{\tt
  arXiv:https://www.science.org/doi/pdf/10.1126/science.abo2196}.
%Type = Article
\bibitem[{Farley et~al.(2020)Farley, Williford, Stack, Bhartia, Chen, de~la
  Torre, Hand, Goreva, Herd, Hueso et~al.}]{farley2020}
\bibinfo{author}{Farley, K.A.}, \bibinfo{author}{Williford, K.H.},
  \bibinfo{author}{Stack, K.M.}, \bibinfo{author}{Bhartia, R.},
  \bibinfo{author}{Chen, A.}, \bibinfo{author}{de~la Torre, M.},
  \bibinfo{author}{Hand, K.}, \bibinfo{author}{Goreva, Y.},
  \bibinfo{author}{Herd, C.D.}, \bibinfo{author}{Hueso, R.}, et~al.,
  \bibinfo{year}{2020}.
\newblock \bibinfo{title}{Mars 2020 mission overview}.
\newblock \bibinfo{journal}{Space Science Reviews} \bibinfo{volume}{216},
  \bibinfo{pages}{1--41}.
\newblock \URLprefix \url{https://doi.org/10.1007/s11214-020-00762-y}.
%Type = Inproceedings
\bibitem[{Forni et~al.(2024)Forni, Bedford, Royer, Liu, Wiens, Dehouck, Meslin,
  Udry, Beyssac, Gabriel, Beck, Gasnault, Quantin-Nataf, Johnson, Schr{\"o}der,
  Pilleri, Debaille, Manelsy, Clark, Cousin, Maurice and Clegg}]{forni_2024}
\bibinfo{author}{Forni, O.}, \bibinfo{author}{Bedford, C.C.},
  \bibinfo{author}{Royer, C.}, \bibinfo{author}{Liu, Y.},
  \bibinfo{author}{Wiens, R.C.}, \bibinfo{author}{Dehouck, E.},
  \bibinfo{author}{Meslin, P.Y.}, \bibinfo{author}{Udry, A.},
  \bibinfo{author}{Beyssac, O.}, \bibinfo{author}{Gabriel, T.S.},
  \bibinfo{author}{Beck, P.}, \bibinfo{author}{Gasnault, O.},
  \bibinfo{author}{Quantin-Nataf, C.}, \bibinfo{author}{Johnson, J.R.},
  \bibinfo{author}{Schr{\"o}der, S.}, \bibinfo{author}{Pilleri, P.},
  \bibinfo{author}{Debaille, V.}, \bibinfo{author}{Manelsy, H.T.},
  \bibinfo{author}{Clark, B.C.}, \bibinfo{author}{Cousin, A.},
  \bibinfo{author}{Maurice, S.}, \bibinfo{author}{Clegg, S.M.},
  \bibinfo{year}{2024}.
\newblock \bibinfo{title}{Nickel-copper deposits on {Mars}? discovery of
  ore-grade abundances, and implications on formation and alteration}, in:
  \bibinfo{booktitle}{{Lunar and Planetary Science Conference}},
  \bibinfo{address}{The Woodlands (Texas), United States}.
\newblock \URLprefix \url{https://hal.science/hal-04446048}.
%Type = Article
\bibitem[{Francis et~al.(2017)Francis, Estlin, Doran, Johnstone, Gaines, Verma,
  Burl, Frydenvang, Monta{\~n}o, Wiens et~al.}]{francis2017}
\bibinfo{author}{Francis, R.}, \bibinfo{author}{Estlin, T.},
  \bibinfo{author}{Doran, G.}, \bibinfo{author}{Johnstone, S.},
  \bibinfo{author}{Gaines, D.}, \bibinfo{author}{Verma, V.},
  \bibinfo{author}{Burl, M.}, \bibinfo{author}{Frydenvang, J.},
  \bibinfo{author}{Monta{\~n}o, S.}, \bibinfo{author}{Wiens, R.}, et~al.,
  \bibinfo{year}{2017}.
\newblock \bibinfo{title}{{AEGIS} autonomous targeting for {ChemCam} on {Mars
  Science Laboratory}: Deployment and results of initial science team use}.
\newblock \bibinfo{journal}{Science Robotics} \bibinfo{volume}{2},
  \bibinfo{pages}{eaan4582}.
\newblock \URLprefix
  \url{https://www.science.org/doi/abs/10.1126/scirobotics.aan4582}.
%Type = Article
\bibitem[{Goetz et~al.(2023)Goetz, Dehouck, Gasda, Johnson, Meslin, Lanza,
  Wiens, Rapin, Frydenvang, Payré and Gasnault}]{goetz_2023}
\bibinfo{author}{Goetz, W.}, \bibinfo{author}{Dehouck, E.},
  \bibinfo{author}{Gasda, P.J.}, \bibinfo{author}{Johnson, J.R.},
  \bibinfo{author}{Meslin, P.Y.}, \bibinfo{author}{Lanza, N.L.},
  \bibinfo{author}{Wiens, R.C.}, \bibinfo{author}{Rapin, W.},
  \bibinfo{author}{Frydenvang, J.}, \bibinfo{author}{Payré, V.},
  \bibinfo{author}{Gasnault, O.}, \bibinfo{year}{2023}.
\newblock \bibinfo{title}{Detection of copper by the {ChemCam} instrument along
  {Curiosity}'s traverse in {Gale} crater, {Mars}: Elevated abundances in {Glen
  Torridon}}.
\newblock \bibinfo{journal}{Journal of Geophysical Research: Planets}
  \bibinfo{volume}{128}, \bibinfo{pages}{e2021JE007101}.
\newblock \URLprefix
  \url{https://agupubs.onlinelibrary.wiley.com/doi/abs/10.1029/2021JE007101},
  \DOIprefix\doi{https://doi.org/10.1029/2021JE007101},
  \href{http://arxiv.org/abs/https://agupubs.onlinelibrary.wiley.com/doi/pdf/10.1029/2021JE007101}{\tt
  arXiv:https://agupubs.onlinelibrary.wiley.com/doi/pdf/10.1029/2021JE007101}.
  \bibinfo{note}{e2021JE007101 2021JE007101}.
%Type = Article
\bibitem[{Goodrich et~al.(2003)Goodrich, Herd and Taylor}]{goodrich_2003}
\bibinfo{author}{Goodrich, C.A.}, \bibinfo{author}{Herd, C.D.K.},
  \bibinfo{author}{Taylor, L.A.}, \bibinfo{year}{2003}.
\newblock \bibinfo{title}{Spinels and oxygen fugacity in olivine-phyric and
  lherzolitic shergottites}.
\newblock \bibinfo{journal}{Meteoritics \& Planetary Science}
  \bibinfo{volume}{38}, \bibinfo{pages}{1773--1792}.
\newblock \URLprefix
  \url{https://onlinelibrary.wiley.com/doi/abs/10.1111/j.1945-5100.2003.tb00014.x},
  \DOIprefix\doi{https://doi.org/10.1111/j.1945-5100.2003.tb00014.x},
  \href{http://arxiv.org/abs/https://onlinelibrary.wiley.com/doi/pdf/10.1111/j.1945-5100.2003.tb00014.x}{\tt
  arXiv:https://onlinelibrary.wiley.com/doi/pdf/10.1111/j.1945-5100.2003.tb00014.x}.
%Type = Misc
\bibitem[{Heirwegh et~al.(2024)Heirwegh, Elam, Liu, Das, Hummel, Naylor, Wade,
  Allwood, Hurowitz, Armstrong, Bacop, O'Neil, Sinclair, Sondheim, Denise,
  Lawson, Rosas, Kawamura, Au, Kitiyakara, Foote, Romero, Anderson, Rossman and
  au2}]{heirwegh_2024}
\bibinfo{author}{Heirwegh, C.M.}, \bibinfo{author}{Elam, W.T.},
  \bibinfo{author}{Liu, Y.}, \bibinfo{author}{Das, A.},
  \bibinfo{author}{Hummel, C.}, \bibinfo{author}{Naylor, B.},
  \bibinfo{author}{Wade, L.A.}, \bibinfo{author}{Allwood, A.C.},
  \bibinfo{author}{Hurowitz, J.A.}, \bibinfo{author}{Armstrong, L.G.},
  \bibinfo{author}{Bacop, N.}, \bibinfo{author}{O'Neil, L.P.},
  \bibinfo{author}{Sinclair, K.P.}, \bibinfo{author}{Sondheim, M.E.},
  \bibinfo{author}{Denise, R.W.}, \bibinfo{author}{Lawson, P.R.},
  \bibinfo{author}{Rosas, R.}, \bibinfo{author}{Kawamura, J.H.},
  \bibinfo{author}{Au, M.H.}, \bibinfo{author}{Kitiyakara, A.},
  \bibinfo{author}{Foote, M.C.}, \bibinfo{author}{Romero, R.A.},
  \bibinfo{author}{Anderson, M.S.}, \bibinfo{author}{Rossman, G.R.},
  \bibinfo{author}{au2, B.C.C.I.}, \bibinfo{year}{2024}.
\newblock \bibinfo{title}{Pre-flight calibration of {PIXL} for {X-ray}
  fluorescence elemental quantification}.
\newblock \URLprefix \url{https://arxiv.org/abs/2402.01544},
  \href{http://arxiv.org/abs/2402.01544}{\tt arXiv:2402.01544}.
%Type = Article
\bibitem[{Heirwegh et~al.(2022)Heirwegh, Elam, O’Neil, Sinclair and
  Das}]{heirwegh_2022}
\bibinfo{author}{Heirwegh, C.M.}, \bibinfo{author}{Elam, W.T.},
  \bibinfo{author}{O’Neil, L.P.}, \bibinfo{author}{Sinclair, K.P.},
  \bibinfo{author}{Das, A.}, \bibinfo{year}{2022}.
\newblock \bibinfo{title}{The focused beam {X-ray} fluorescence elemental
  quantification software package {PIQUANT}}.
\newblock \bibinfo{journal}{Spectrochim. Acta B} \bibinfo{volume}{196},
  \bibinfo{pages}{106520}.
\newblock \URLprefix \url{https://doi.org/10.1016/j.sab.2022.106520}.
%Type = Inbook
\bibitem[{Herd et~al.(2018)Herd, Moser, Tait, Darling, Shaulis and
  McCoy}]{herd_2017}
\bibinfo{author}{Herd, C.D.K.}, \bibinfo{author}{Moser, D.E.},
  \bibinfo{author}{Tait, K.}, \bibinfo{author}{Darling, J.R.},
  \bibinfo{author}{Shaulis, B.J.}, \bibinfo{author}{McCoy, T.J.},
  \bibinfo{year}{2018}.
\newblock \bibinfo{title}{Crystallization of Baddeleyite in Basaltic Rocks from
  {Mars}, and Comparisons with the {Earth, Moon, and Vesta}}.
  \bibinfo{publisher}{American Geophysical Union (AGU)}.
  chapter~\bibinfo{chapter}{6}.
\newblock pp. \bibinfo{pages}{137--166}.
\newblock \URLprefix
  \url{https://agupubs.onlinelibrary.wiley.com/doi/abs/10.1002/9781119227250.ch6},
  \DOIprefix\doi{https://doi.org/10.1002/9781119227250.ch6},
  \href{http://arxiv.org/abs/https://agupubs.onlinelibrary.wiley.com/doi/pdf/10.1002/9781119227250.ch6}{\tt
  arXiv:https://agupubs.onlinelibrary.wiley.com/doi/pdf/10.1002/9781119227250.ch6}.
%Type = Inproceedings
\bibitem[{Lanza et~al.(2015)Lanza, Wiens, Arvidson, Clark, Fischer, Gellert,
  Grotzinger, Hurowitz, McLennan, Morris, Rice, Bell, Berger, Blaney, Blank,
  Bridges, Calef, Campbell, Clegg, Cousin, Edgett, Fabre, Fisk, Forni,
  Frydenvang, Hardy, Hardgrove, Johnson, Kah, Lasue, {Le Mouelic}, Malin,
  Mangold, Martin-Torres, Maurice, McBride, Ming, Newsom, Schroder, Thompson,
  Treiman, VanBommel, Vaniman and Zorzano}]{lanza_2015}
\bibinfo{author}{Lanza, N.}, \bibinfo{author}{Wiens, R.},
  \bibinfo{author}{Arvidson, R.}, \bibinfo{author}{Clark, B.},
  \bibinfo{author}{Fischer, W.}, \bibinfo{author}{Gellert, R.},
  \bibinfo{author}{Grotzinger, J.}, \bibinfo{author}{Hurowitz, J.},
  \bibinfo{author}{McLennan, S.}, \bibinfo{author}{Morris, R.},
  \bibinfo{author}{Rice, M.}, \bibinfo{author}{Bell, J.},
  \bibinfo{author}{Berger, J.}, \bibinfo{author}{Blaney, D.},
  \bibinfo{author}{Blank, J.}, \bibinfo{author}{Bridges, N.},
  \bibinfo{author}{Calef, F.}, \bibinfo{author}{Campbell, J.},
  \bibinfo{author}{Clegg, S.}, \bibinfo{author}{Cousin, A.},
  \bibinfo{author}{Edgett, K.}, \bibinfo{author}{Fabre, C.},
  \bibinfo{author}{Fisk, M.}, \bibinfo{author}{Forni, O.},
  \bibinfo{author}{Frydenvang, J.}, \bibinfo{author}{Hardy, K.},
  \bibinfo{author}{Hardgrove, C.}, \bibinfo{author}{Johnson, J.},
  \bibinfo{author}{Kah, L.}, \bibinfo{author}{Lasue, J.}, \bibinfo{author}{{Le
  Mouelic}, S.}, \bibinfo{author}{Malin, M.}, \bibinfo{author}{Mangold, N.},
  \bibinfo{author}{Martin-Torres, J.}, \bibinfo{author}{Maurice, S.},
  \bibinfo{author}{McBride, M.}, \bibinfo{author}{Ming, D.},
  \bibinfo{author}{Newsom, H.}, \bibinfo{author}{Schroder, S.},
  \bibinfo{author}{Thompson, L.}, \bibinfo{author}{Treiman, A.},
  \bibinfo{author}{VanBommel, S.}, \bibinfo{author}{Vaniman, D.},
  \bibinfo{author}{Zorzano, M.}, \bibinfo{year}{2015}.
\newblock \bibinfo{title}{Oxidation of manganese at {Kimberley}, {Gale} crater:
  More free oxygen in {Mars}' past?}
\newblock \URLprefix \url{https://www.hou.usra.edu/meetings/lpsc2015/}.
  \bibinfo{note}{lPI Contribution No. 1832, p.2893; 46th Lunar and Planetary
  Science Conference ; Conference date: 16-03-2015 Through 20-03-2015}.
%Type = Article
\bibitem[{Lanza et~al.(2014)Lanza, Fischer, Wiens, Grotzinger, Ollila, Cousin,
  Anderson, Clark, Gellert, Mangold, Maurice, Le~Mouélic, Nachon, Schmidt,
  Berger, Clegg, Forni, Hardgrove, Melikechi, Newsom and Sautter}]{lanza_2014}
\bibinfo{author}{Lanza, N.L.}, \bibinfo{author}{Fischer, W.W.},
  \bibinfo{author}{Wiens, R.C.}, \bibinfo{author}{Grotzinger, J.},
  \bibinfo{author}{Ollila, A.M.}, \bibinfo{author}{Cousin, A.},
  \bibinfo{author}{Anderson, R.B.}, \bibinfo{author}{Clark, B.C.},
  \bibinfo{author}{Gellert, R.}, \bibinfo{author}{Mangold, N.},
  \bibinfo{author}{Maurice, S.}, \bibinfo{author}{Le~Mouélic, S.},
  \bibinfo{author}{Nachon, M.}, \bibinfo{author}{Schmidt, M.},
  \bibinfo{author}{Berger, J.}, \bibinfo{author}{Clegg, S.M.},
  \bibinfo{author}{Forni, O.}, \bibinfo{author}{Hardgrove, C.},
  \bibinfo{author}{Melikechi, N.}, \bibinfo{author}{Newsom, H.E.},
  \bibinfo{author}{Sautter, V.}, \bibinfo{year}{2014}.
\newblock \bibinfo{title}{High manganese concentrations in rocks at {Gale}
  crater, {Mars}}.
\newblock \bibinfo{journal}{Geophysical Research Letters} \bibinfo{volume}{41},
  \bibinfo{pages}{5755--5763}.
\newblock \URLprefix
  \url{https://agupubs.onlinelibrary.wiley.com/doi/abs/10.1002/2014GL060329},
  \DOIprefix\doi{https://doi.org/10.1002/2014GL060329},
  \href{http://arxiv.org/abs/https://agupubs.onlinelibrary.wiley.com/doi/pdf/10.1002/2014GL060329}{\tt
  arXiv:https://agupubs.onlinelibrary.wiley.com/doi/pdf/10.1002/2014GL060329}.
%Type = Article
\bibitem[{Liu et~al.(2022)Liu, Tice, Schmidt, Treiman, Kizovski, Hurowitz,
  Allwood, Henneke, Pedersen, VanBommel, Jones, Knight, Orenstein, Clark, Elam,
  Heirwegh, Barber, Beegle, Benzerara, Bernard, Beyssac, Bosak, Brown,
  Cardarelli, Catling, Christian, Cloutis, Cohen, Davidoff, Fairén, Farley,
  Flannery, Galvin, Grotzinger, Gupta, Hall, Herd, Hickman-Lewis, Hodyss,
  Horgan, Johnson, Jørgensen, Kah, Maki, Mandon, Mangold, McCubbin, McLennan,
  Moore, Nachon, Nemere, Nothdurft, Núñez, O’Neil, Quantin-Nataf, Sautter,
  Shuster, Siebach, Simon, Sinclair, Stack, Steele, Tarnas, Tosca, Uckert,
  Udry, Wade, Weiss, Wiens, Williford and Zorzano}]{liu_2022}
\bibinfo{author}{Liu, Y.}, \bibinfo{author}{Tice, M.M.},
  \bibinfo{author}{Schmidt, M.E.}, \bibinfo{author}{Treiman, A.H.},
  \bibinfo{author}{Kizovski, T.V.}, \bibinfo{author}{Hurowitz, J.A.},
  \bibinfo{author}{Allwood, A.C.}, \bibinfo{author}{Henneke, J.},
  \bibinfo{author}{Pedersen, D.A.K.}, \bibinfo{author}{VanBommel, S.J.},
  \bibinfo{author}{Jones, M.W.M.}, \bibinfo{author}{Knight, A.L.},
  \bibinfo{author}{Orenstein, B.J.}, \bibinfo{author}{Clark, B.C.},
  \bibinfo{author}{Elam, W.T.}, \bibinfo{author}{Heirwegh, C.M.},
  \bibinfo{author}{Barber, T.}, \bibinfo{author}{Beegle, L.W.},
  \bibinfo{author}{Benzerara, K.}, \bibinfo{author}{Bernard, S.},
  \bibinfo{author}{Beyssac, O.}, \bibinfo{author}{Bosak, T.},
  \bibinfo{author}{Brown, A.J.}, \bibinfo{author}{Cardarelli, E.L.},
  \bibinfo{author}{Catling, D.C.}, \bibinfo{author}{Christian, J.R.},
  \bibinfo{author}{Cloutis, E.A.}, \bibinfo{author}{Cohen, B.A.},
  \bibinfo{author}{Davidoff, S.}, \bibinfo{author}{Fairén, A.G.},
  \bibinfo{author}{Farley, K.A.}, \bibinfo{author}{Flannery, D.T.},
  \bibinfo{author}{Galvin, A.}, \bibinfo{author}{Grotzinger, J.P.},
  \bibinfo{author}{Gupta, S.}, \bibinfo{author}{Hall, J.},
  \bibinfo{author}{Herd, C.D.K.}, \bibinfo{author}{Hickman-Lewis, K.},
  \bibinfo{author}{Hodyss, R.P.}, \bibinfo{author}{Horgan, B.H.N.},
  \bibinfo{author}{Johnson, J.R.}, \bibinfo{author}{Jørgensen, J.L.},
  \bibinfo{author}{Kah, L.C.}, \bibinfo{author}{Maki, J.N.},
  \bibinfo{author}{Mandon, L.}, \bibinfo{author}{Mangold, N.},
  \bibinfo{author}{McCubbin, F.M.}, \bibinfo{author}{McLennan, S.M.},
  \bibinfo{author}{Moore, K.}, \bibinfo{author}{Nachon, M.},
  \bibinfo{author}{Nemere, P.}, \bibinfo{author}{Nothdurft, L.D.},
  \bibinfo{author}{Núñez, J.I.}, \bibinfo{author}{O’Neil, L.},
  \bibinfo{author}{Quantin-Nataf, C.M.}, \bibinfo{author}{Sautter, V.},
  \bibinfo{author}{Shuster, D.L.}, \bibinfo{author}{Siebach, K.L.},
  \bibinfo{author}{Simon, J.I.}, \bibinfo{author}{Sinclair, K.P.},
  \bibinfo{author}{Stack, K.M.}, \bibinfo{author}{Steele, A.},
  \bibinfo{author}{Tarnas, J.D.}, \bibinfo{author}{Tosca, N.J.},
  \bibinfo{author}{Uckert, K.}, \bibinfo{author}{Udry, A.},
  \bibinfo{author}{Wade, L.A.}, \bibinfo{author}{Weiss, B.P.},
  \bibinfo{author}{Wiens, R.C.}, \bibinfo{author}{Williford, K.H.},
  \bibinfo{author}{Zorzano, M.P.}, \bibinfo{year}{2022}.
\newblock \bibinfo{title}{An olivine cumulate outcrop on the floor of jezero
  crater, mars}.
\newblock \bibinfo{journal}{Science} \bibinfo{volume}{377},
  \bibinfo{pages}{1513--1519}.
\newblock \URLprefix
  \url{https://www.science.org/doi/abs/10.1126/science.abo2756},
  \DOIprefix\doi{10.1126/science.abo2756},
  \href{http://arxiv.org/abs/https://www.science.org/doi/pdf/10.1126/science.abo2756}{\tt
  arXiv:https://www.science.org/doi/pdf/10.1126/science.abo2756}.
%Type = Article
\bibitem[{McCubbin et~al.(2016)McCubbin, Boyce, Srinivasan, Santos, Elardo,
  Filiberto, Steele and Shearer}]{mccubbin_2016}
\bibinfo{author}{McCubbin, F.M.}, \bibinfo{author}{Boyce, J.W.},
  \bibinfo{author}{Srinivasan, P.}, \bibinfo{author}{Santos, A.R.},
  \bibinfo{author}{Elardo, S.M.}, \bibinfo{author}{Filiberto, J.},
  \bibinfo{author}{Steele, A.}, \bibinfo{author}{Shearer, C.K.},
  \bibinfo{year}{2016}.
\newblock \bibinfo{title}{Heterogeneous distribution of {H$_2$O} in the martian
  interior: Implications for the abundance of {H$_2$O} in depleted and enriched
  mantle sources}.
\newblock \bibinfo{journal}{Meteoritics \& Planetary Science}
  \bibinfo{volume}{51}, \bibinfo{pages}{2036--2060}.
\newblock \URLprefix
  \url{https://onlinelibrary.wiley.com/doi/abs/10.1111/maps.12639},
  \DOIprefix\doi{https://doi.org/10.1111/maps.12639},
  \href{http://arxiv.org/abs/https://onlinelibrary.wiley.com/doi/pdf/10.1111/maps.12639}{\tt
  arXiv:https://onlinelibrary.wiley.com/doi/pdf/10.1111/maps.12639}.
%Type = Article
\bibitem[{Moeller et~al.(2021)Moeller, Jandura, Rosette, Robinson, Samuels,
  Silverman, Brown, Duffy, Yazzie, Jens, Brockie, White, Goreva, Zorn, Okon,
  Lin, Frost, Collins, Williams and et~al.}]{moeller_2021}
\bibinfo{author}{Moeller, R.}, \bibinfo{author}{Jandura, L.},
  \bibinfo{author}{Rosette, K.}, \bibinfo{author}{Robinson, M.},
  \bibinfo{author}{Samuels, J.}, \bibinfo{author}{Silverman, M.},
  \bibinfo{author}{Brown, K.}, \bibinfo{author}{Duffy, E.},
  \bibinfo{author}{Yazzie, A.}, \bibinfo{author}{Jens, E.},
  \bibinfo{author}{Brockie, I.}, \bibinfo{author}{White, L.},
  \bibinfo{author}{Goreva, Y.}, \bibinfo{author}{Zorn, T.},
  \bibinfo{author}{Okon, A.}, \bibinfo{author}{Lin, J.},
  \bibinfo{author}{Frost, M.}, \bibinfo{author}{Collins, C.},
  \bibinfo{author}{Williams, J.}, \bibinfo{author}{et~al.},
  \bibinfo{year}{2021}.
\newblock \bibinfo{title}{The sampling and caching subsystem {(SCS)} for the
  scientific exploration of {Jezero} crater by the {Mars 2020 Perseverance}
  rover}.
\newblock \bibinfo{journal}{Space Sci. Rev.} \bibinfo{volume}{217}.
\newblock \URLprefix \url{https://doi.org/10.1007/ s11214-020-00783-7}.
%Type = Article
\bibitem[{Nemere et~al.(2024)Nemere, Barber, Stonebraker, Fedell, Galvin and
  Davidoff}]{nemere_2024}
\bibinfo{author}{Nemere, P.}, \bibinfo{author}{Barber, T.},
  \bibinfo{author}{Stonebraker, R.}, \bibinfo{author}{Fedell, S.M.},
  \bibinfo{author}{Galvin, A.}, \bibinfo{author}{Davidoff, S.},
  \bibinfo{year}{2024}.
\newblock \bibinfo{title}{{PIXLISE} core}.
\newblock \bibinfo{journal}{Zenodo} \URLprefix
  \url{https://doi.org/10.5281/zenodo.10183256}.
%Type = Article
\bibitem[{Papike et~al.(2009)Papike, Karner, Shearer and Burger}]{papike_2009}
\bibinfo{author}{Papike, J.}, \bibinfo{author}{Karner, J.},
  \bibinfo{author}{Shearer, C.}, \bibinfo{author}{Burger, P.},
  \bibinfo{year}{2009}.
\newblock \bibinfo{title}{Silicate mineralogy of martian meteorites}.
\newblock \bibinfo{journal}{Geochimica et Cosmochimica Acta}
  \bibinfo{volume}{73}, \bibinfo{pages}{7443--7485}.
\newblock \URLprefix
  \url{https://www.sciencedirect.com/science/article/pii/S0016703709005651},
  \DOIprefix\doi{https://doi.org/10.1016/j.gca.2009.09.008}.
%Type = Article
\bibitem[{Patel et~al.(2015)Patel, Percivalle, Ritson, Duffy and
  Sutherland}]{patel_2015}
\bibinfo{author}{Patel, B.H.}, \bibinfo{author}{Percivalle, C.},
  \bibinfo{author}{Ritson, D.J.}, \bibinfo{author}{Duffy, C.D.},
  \bibinfo{author}{Sutherland, J.D.}, \bibinfo{year}{2015}.
\newblock \bibinfo{title}{Common origins of {RNA}, protein and lipid precursors
  in a cyanosulfidic protometabolism}.
\newblock \bibinfo{journal}{Nature Chemistry} \bibinfo{volume}{7},
  \bibinfo{pages}{301--307}.
\newblock \DOIprefix\doi{https://doi.org/10.1038/nchem.2202}.
%Type = Article
\bibitem[{Payré et~al.(2019)Payré, Fabre, Sautter, Cousin, Mangold, Deit,
  Forni, Goetz, Wiens, Gasnault, Meslin, Lasue, Rapin, Clark, Nachon, Lanza and
  Maurice}]{payre_2019}
\bibinfo{author}{Payré, V.}, \bibinfo{author}{Fabre, C.},
  \bibinfo{author}{Sautter, V.}, \bibinfo{author}{Cousin, A.},
  \bibinfo{author}{Mangold, N.}, \bibinfo{author}{Deit, L.L.},
  \bibinfo{author}{Forni, O.}, \bibinfo{author}{Goetz, W.},
  \bibinfo{author}{Wiens, R.C.}, \bibinfo{author}{Gasnault, O.},
  \bibinfo{author}{Meslin, P.Y.}, \bibinfo{author}{Lasue, J.},
  \bibinfo{author}{Rapin, W.}, \bibinfo{author}{Clark, B.},
  \bibinfo{author}{Nachon, M.}, \bibinfo{author}{Lanza, N.L.},
  \bibinfo{author}{Maurice, S.}, \bibinfo{year}{2019}.
\newblock \bibinfo{title}{Copper enrichments in the {Kimberley} formation in
  {Gale} crater, {Mars}: Evidence for a {Cu} deposit at the source}.
\newblock \bibinfo{journal}{Icarus} \bibinfo{volume}{321},
  \bibinfo{pages}{736--751}.
\newblock \URLprefix
  \url{https://www.sciencedirect.com/science/article/pii/S0019103518304779},
  \DOIprefix\doi{https://doi.org/10.1016/j.icarus.2018.12.015}.
%Type = Article
\bibitem[{Pedregosa et~al.(2011)Pedregosa, Varoquaux, Gramfort, Michel,
  Thirion, Grisel, Blondel, Prettenhofer, Weiss, Dubourg, Vanderplas, Passos,
  Cournapeau, Brucher, Perrot and {{\'E}}douard Duchesnay}]{pedregosa_2011}
\bibinfo{author}{Pedregosa, F.}, \bibinfo{author}{Varoquaux, G.},
  \bibinfo{author}{Gramfort, A.}, \bibinfo{author}{Michel, V.},
  \bibinfo{author}{Thirion, B.}, \bibinfo{author}{Grisel, O.},
  \bibinfo{author}{Blondel, M.}, \bibinfo{author}{Prettenhofer, P.},
  \bibinfo{author}{Weiss, R.}, \bibinfo{author}{Dubourg, V.},
  \bibinfo{author}{Vanderplas, J.}, \bibinfo{author}{Passos, A.},
  \bibinfo{author}{Cournapeau, D.}, \bibinfo{author}{Brucher, M.},
  \bibinfo{author}{Perrot, M.}, \bibinfo{author}{{{\'E}}douard Duchesnay},
  \bibinfo{year}{2011}.
\newblock \bibinfo{title}{Scikit-learn: Machine learning in python}.
\newblock \bibinfo{journal}{Journal of Machine Learning Research}
  \bibinfo{volume}{12}, \bibinfo{pages}{2825--2830}.
\newblock \URLprefix \url{http://jmlr.org/papers/v12/pedregosa11a.html}.
%Type = Book
\bibitem[{Russell and Norvig(2020)}]{Russel2021}
\bibinfo{author}{Russell, S.J.}, \bibinfo{author}{Norvig, P.},
  \bibinfo{year}{2020}.
\newblock \bibinfo{title}{Artificial Intelligence: {A} Modern Approach (4th
  Edition)}.
\newblock \bibinfo{publisher}{Pearson}.
\newblock \URLprefix \url{http://aima.cs.berkeley.edu/}.
%Type = Article
\bibitem[{Scheller et~al.(2024 submitted)Scheller, Bosak, McCubbin, Williford,
  Siljeström, Jakubek, Eckley, Morris, Bykov, Kizovski, Asher, Berger, Bower,
  Cardarelli, Ehlmann, Fornaro, Fox, Haney, Hand, Roppel, Sharma, Steele,
  Uckert, Yanchilina, Beyssac, Farley, Henneke, Heirwegh, Pedersen, Liu,
  Schmidt, Sephton, Shuster and Weiss}]{scheller_2024}
\bibinfo{author}{Scheller, E.}, \bibinfo{author}{Bosak, T.},
  \bibinfo{author}{McCubbin, F.}, \bibinfo{author}{Williford, K.},
  \bibinfo{author}{Siljeström, S.}, \bibinfo{author}{Jakubek, R.},
  \bibinfo{author}{Eckley, S.}, \bibinfo{author}{Morris, R.},
  \bibinfo{author}{Bykov, S.}, \bibinfo{author}{Kizovski, T.},
  \bibinfo{author}{Asher, S.}, \bibinfo{author}{Berger, E.},
  \bibinfo{author}{Bower, D.}, \bibinfo{author}{Cardarelli, E.},
  \bibinfo{author}{Ehlmann, B.}, \bibinfo{author}{Fornaro, T.},
  \bibinfo{author}{Fox, A.}, \bibinfo{author}{Haney, N.},
  \bibinfo{author}{Hand, K.}, \bibinfo{author}{Roppel, R.},
  \bibinfo{author}{Sharma, S.}, \bibinfo{author}{Steele, A.},
  \bibinfo{author}{Uckert, K.}, \bibinfo{author}{Yanchilina, A.},
  \bibinfo{author}{Beyssac, O.}, \bibinfo{author}{Farley, K.},
  \bibinfo{author}{Henneke, J.}, \bibinfo{author}{Heirwegh, C.},
  \bibinfo{author}{Pedersen, D.}, \bibinfo{author}{Liu, Y.},
  \bibinfo{author}{Schmidt, M.}, \bibinfo{author}{Sephton, M.},
  \bibinfo{author}{Shuster, D.}, \bibinfo{author}{Weiss, B.},
  \bibinfo{year}{2024 submitted}.
\newblock \bibinfo{title}{Inorganic interpretation of luminescent materials
  encountered by the {Perseverance} rover on {Mars}}.
\newblock \bibinfo{journal}{Science Advances} .
%Type = Article
\bibitem[{Sun et~al.(2023)Sun, Hand, Stack, Farley, Simon, Newman, Sharma, Liu,
  Wiens, Williams, Tosca, Alwmark, Beyssac, Brown, Calef, Cardarelli, Clavé,
  Cohen, Corpolongo, Czaja, Del~Sesto, Fairen, Fornaro, Fouchet, Garczynski,
  Gupta, Herd, Hickman-Lewis, Horgan, Johnson, Kinch, Kizovski, Kronyak, Lange,
  Mandon, Milkovich, Moeller, Núñez, Paar, Pyrzak, Quantin-Nataf, Shuster,
  Siljestrom, Steele, Tice, Toupet, Udry, Vaughan and Wogsland}]{sun_2023}
\bibinfo{author}{Sun, V.Z.}, \bibinfo{author}{Hand, K.P.},
  \bibinfo{author}{Stack, K.M.}, \bibinfo{author}{Farley, K.A.},
  \bibinfo{author}{Simon, J.I.}, \bibinfo{author}{Newman, C.},
  \bibinfo{author}{Sharma, S.}, \bibinfo{author}{Liu, Y.},
  \bibinfo{author}{Wiens, R.C.}, \bibinfo{author}{Williams, A.J.},
  \bibinfo{author}{Tosca, N.}, \bibinfo{author}{Alwmark, S.},
  \bibinfo{author}{Beyssac, O.}, \bibinfo{author}{Brown, A.},
  \bibinfo{author}{Calef, F.}, \bibinfo{author}{Cardarelli, E.L.},
  \bibinfo{author}{Clavé, E.}, \bibinfo{author}{Cohen, B.},
  \bibinfo{author}{Corpolongo, A.}, \bibinfo{author}{Czaja, A.D.},
  \bibinfo{author}{Del~Sesto, T.}, \bibinfo{author}{Fairen, A.},
  \bibinfo{author}{Fornaro, T.}, \bibinfo{author}{Fouchet, T.},
  \bibinfo{author}{Garczynski, B.}, \bibinfo{author}{Gupta, S.},
  \bibinfo{author}{Herd, C.D.K.}, \bibinfo{author}{Hickman-Lewis, K.},
  \bibinfo{author}{Horgan, B.}, \bibinfo{author}{Johnson, J.},
  \bibinfo{author}{Kinch, K.}, \bibinfo{author}{Kizovski, T.},
  \bibinfo{author}{Kronyak, R.}, \bibinfo{author}{Lange, R.},
  \bibinfo{author}{Mandon, L.}, \bibinfo{author}{Milkovich, S.},
  \bibinfo{author}{Moeller, R.}, \bibinfo{author}{Núñez, J.},
  \bibinfo{author}{Paar, G.}, \bibinfo{author}{Pyrzak, G.},
  \bibinfo{author}{Quantin-Nataf, C.}, \bibinfo{author}{Shuster, D.L.},
  \bibinfo{author}{Siljestrom, S.}, \bibinfo{author}{Steele, A.},
  \bibinfo{author}{Tice, M.}, \bibinfo{author}{Toupet, O.},
  \bibinfo{author}{Udry, A.}, \bibinfo{author}{Vaughan, A.},
  \bibinfo{author}{Wogsland, B.}, \bibinfo{year}{2023}.
\newblock \bibinfo{title}{Overview and results from the {Mars 2020
  Perseverance} rover's first science campaign on the {Jezero} crater floor}.
\newblock \bibinfo{journal}{Journal of Geophysical Research: Planets}
  \bibinfo{volume}{128}, \bibinfo{pages}{e2022JE007613}.
\newblock \URLprefix
  \url{https://agupubs.onlinelibrary.wiley.com/doi/abs/10.1029/2022JE007613},
  \DOIprefix\doi{https://doi.org/10.1029/2022JE007613},
  \href{http://arxiv.org/abs/https://agupubs.onlinelibrary.wiley.com/doi/pdf/10.1029/2022JE007613}{\tt
  arXiv:https://agupubs.onlinelibrary.wiley.com/doi/pdf/10.1029/2022JE007613}.
  \bibinfo{note}{e2022JE007613 2022JE007613}.
%Type = Article
\bibitem[{Sutherland(2016)}]{sutherland_2016}
\bibinfo{author}{Sutherland, J.D.}, \bibinfo{year}{2016}.
\newblock \bibinfo{title}{The origin of life—out of the blue}.
\newblock \bibinfo{journal}{Angewandte Chemie International Edition}
  \bibinfo{volume}{55}, \bibinfo{pages}{104--121}.
\newblock \URLprefix
  \url{https://onlinelibrary.wiley.com/doi/abs/10.1002/anie.201506585},
  \DOIprefix\doi{https://doi.org/10.1002/anie.201506585},
  \href{http://arxiv.org/abs/https://onlinelibrary.wiley.com/doi/pdf/10.1002/anie.201506585}{\tt
  arXiv:https://onlinelibrary.wiley.com/doi/pdf/10.1002/anie.201506585}.
%Type = Article
\bibitem[{Thompson et~al.(2015)Thompson, Flannery, Kiran, Allwood, Bue, Clark,
  Elam, Estlin, Hodyss, Hurowitz, Liu and Wade}]{Thompson2015}
\bibinfo{author}{Thompson, D.R.}, \bibinfo{author}{Flannery, D.T.},
  \bibinfo{author}{Kiran, R.A.}, \bibinfo{author}{Allwood, A.C.},
  \bibinfo{author}{Bue, B.D.}, \bibinfo{author}{Clark, B.},
  \bibinfo{author}{Elam, W.T.}, \bibinfo{author}{Estlin, T.},
  \bibinfo{author}{Hodyss, R.}, \bibinfo{author}{Hurowitz, J.A.},
  \bibinfo{author}{Liu, Y.}, \bibinfo{author}{Wade, L.}, \bibinfo{year}{2015}.
\newblock \bibinfo{title}{Automating {X-ray} fluorescence analysis for rapid
  astrobiology surveys}.
\newblock \bibinfo{journal}{Astrobiology} \bibinfo{volume}{15},
  \bibinfo{pages}{961--976}.
\newblock \URLprefix \url{https://doi.org/10.1089/ast.2015.1349},
  \DOIprefix\doi{10.1089/ast.2015.1349}.
%Type = Article
\bibitem[{Thompson et~al.(2011)Thompson, Wettergreen and
  Peralta}]{thompson2011}
\bibinfo{author}{Thompson, D.R.}, \bibinfo{author}{Wettergreen, D.S.},
  \bibinfo{author}{Peralta, F.J.C.}, \bibinfo{year}{2011}.
\newblock \bibinfo{title}{Autonomous science during large-scale robotic
  survey}.
\newblock \bibinfo{journal}{Journal of Field Robotics} \bibinfo{volume}{28},
  \bibinfo{pages}{542--564}.
%Type = Article
\bibitem[{Tice et~al.(2022)Tice, Hurowitz, Allwood, Jones, Orenstein, Davidoff,
  Wright, Pedersen, Henneke, Tosca, Moore, Clark, McLennan, Flannery, Steele,
  Brown, Zorzano, Hickman-Lewis, Liu, VanBommel, Schmidt, Kizovski, Treiman,
  O’Neil, Fairén, Shuster, Gupta and Team}]{tice_2022}
\bibinfo{author}{Tice, M.M.}, \bibinfo{author}{Hurowitz, J.A.},
  \bibinfo{author}{Allwood, A.C.}, \bibinfo{author}{Jones, M.W.M.},
  \bibinfo{author}{Orenstein, B.J.}, \bibinfo{author}{Davidoff, S.},
  \bibinfo{author}{Wright, A.P.}, \bibinfo{author}{Pedersen, D.A.},
  \bibinfo{author}{Henneke, J.}, \bibinfo{author}{Tosca, N.J.},
  \bibinfo{author}{Moore, K.R.}, \bibinfo{author}{Clark, B.C.},
  \bibinfo{author}{McLennan, S.M.}, \bibinfo{author}{Flannery, D.T.},
  \bibinfo{author}{Steele, A.}, \bibinfo{author}{Brown, A.J.},
  \bibinfo{author}{Zorzano, M.P.}, \bibinfo{author}{Hickman-Lewis, K.},
  \bibinfo{author}{Liu, Y.}, \bibinfo{author}{VanBommel, S.J.},
  \bibinfo{author}{Schmidt, M.E.}, \bibinfo{author}{Kizovski, T.V.},
  \bibinfo{author}{Treiman, A.H.}, \bibinfo{author}{O’Neil, L.},
  \bibinfo{author}{Fairén, A.G.}, \bibinfo{author}{Shuster, D.L.},
  \bibinfo{author}{Gupta, S.}, \bibinfo{author}{Team, T.P.},
  \bibinfo{year}{2022}.
\newblock \bibinfo{title}{Alteration history of séítah formation rocks
  inferred by pixl x-ray fluorescence, x-ray diffraction, and multispectral
  imaging on mars}.
\newblock \bibinfo{journal}{Science Advances} \bibinfo{volume}{8},
  \bibinfo{pages}{eabp9084}.
\newblock \URLprefix
  \url{https://www.science.org/doi/abs/10.1126/sciadv.abp9084},
  \DOIprefix\doi{10.1126/sciadv.abp9084},
  \href{http://arxiv.org/abs/https://www.science.org/doi/pdf/10.1126/sciadv.abp9084}{\tt
  arXiv:https://www.science.org/doi/pdf/10.1126/sciadv.abp9084}.
%Type = Article
\bibitem[{Treiman et~al.(2023)Treiman, Lanza, VanBommel, Berger, Wiens,
  Bristow, Johnson, Rice, Hart, McAdam, Gasda, Meslin, Yen, Williams, Vasavada,
  Vaniman, Tu, Thorpe, Swanner, Seeger, Schwenzer, Schröder, Rampe, Rapin,
  Ralston, Peretyazhko, Newsom, Morris, Ming, Loche, Le~Mouélic, House, Hazen,
  Grotzinger, Gellert, Gasnault, Fischer, Essunfeld, Downs, Downs, Dehouck,
  Crossey, Cousin, Comellas, Clark, Clark, Chipera, Caravaca, Bridges, Blake
  and Anderson}]{treiman_2023}
\bibinfo{author}{Treiman, A.H.}, \bibinfo{author}{Lanza, N.L.},
  \bibinfo{author}{VanBommel, S.}, \bibinfo{author}{Berger, J.},
  \bibinfo{author}{Wiens, R.}, \bibinfo{author}{Bristow, T.},
  \bibinfo{author}{Johnson, J.}, \bibinfo{author}{Rice, M.},
  \bibinfo{author}{Hart, R.}, \bibinfo{author}{McAdam, A.},
  \bibinfo{author}{Gasda, P.}, \bibinfo{author}{Meslin, P.Y.},
  \bibinfo{author}{Yen, A.}, \bibinfo{author}{Williams, A.J.},
  \bibinfo{author}{Vasavada, A.}, \bibinfo{author}{Vaniman, D.},
  \bibinfo{author}{Tu, V.}, \bibinfo{author}{Thorpe, M.},
  \bibinfo{author}{Swanner, E.D.}, \bibinfo{author}{Seeger, C.},
  \bibinfo{author}{Schwenzer, S.P.}, \bibinfo{author}{Schröder, S.},
  \bibinfo{author}{Rampe, E.}, \bibinfo{author}{Rapin, W.},
  \bibinfo{author}{Ralston, S.J.}, \bibinfo{author}{Peretyazhko, T.},
  \bibinfo{author}{Newsom, H.}, \bibinfo{author}{Morris, R.V.},
  \bibinfo{author}{Ming, D.}, \bibinfo{author}{Loche, M.},
  \bibinfo{author}{Le~Mouélic, S.}, \bibinfo{author}{House, C.},
  \bibinfo{author}{Hazen, R.}, \bibinfo{author}{Grotzinger, J.P.},
  \bibinfo{author}{Gellert, R.}, \bibinfo{author}{Gasnault, O.},
  \bibinfo{author}{Fischer, W.W.}, \bibinfo{author}{Essunfeld, A.},
  \bibinfo{author}{Downs, R.T.}, \bibinfo{author}{Downs, G.W.},
  \bibinfo{author}{Dehouck, E.}, \bibinfo{author}{Crossey, L.J.},
  \bibinfo{author}{Cousin, A.}, \bibinfo{author}{Comellas, J.M.},
  \bibinfo{author}{Clark, J.V.}, \bibinfo{author}{Clark, B.},
  \bibinfo{author}{Chipera, S.}, \bibinfo{author}{Caravaca, G.},
  \bibinfo{author}{Bridges, J.}, \bibinfo{author}{Blake, D.F.},
  \bibinfo{author}{Anderson, R.}, \bibinfo{year}{2023}.
\newblock \bibinfo{title}{Manganese-iron phosphate nodules at the {Groken}
  site, {Gale} crater, {Mars}}.
\newblock \bibinfo{journal}{Minerals} \bibinfo{volume}{13}.
\newblock \URLprefix \url{https://www.mdpi.com/2075-163X/13/9/1122},
  \DOIprefix\doi{10.3390/min13091122}.
%Type = Article
\bibitem[{Udry et~al.(2020)Udry, Howarth, Herd, Day, Lapen and
  Filiberto}]{udry_2020}
\bibinfo{author}{Udry, A.}, \bibinfo{author}{Howarth, G.H.},
  \bibinfo{author}{Herd, C.D.K.}, \bibinfo{author}{Day, J.M.D.},
  \bibinfo{author}{Lapen, T.J.}, \bibinfo{author}{Filiberto, J.},
  \bibinfo{year}{2020}.
\newblock \bibinfo{title}{What martian meteorites reveal about the interior and
  surface of {Mars}}.
\newblock \bibinfo{journal}{Journal of Geophysical Research: Planets}
  \bibinfo{volume}{125}, \bibinfo{pages}{e2020JE006523}.
\newblock \URLprefix
  \url{https://agupubs.onlinelibrary.wiley.com/doi/abs/10.1029/2020JE006523},
  \DOIprefix\doi{https://doi.org/10.1029/2020JE006523},
  \href{http://arxiv.org/abs/https://agupubs.onlinelibrary.wiley.com/doi/pdf/10.1029/2020JE006523}{\tt
  arXiv:https://agupubs.onlinelibrary.wiley.com/doi/pdf/10.1029/2020JE006523}.
  \bibinfo{note}{e2020JE006523 2020JE006523}.
%Type = Book
\bibitem[{{Van Grieken} and Markowicz(2001)}]{VanGrieken2001}
\bibinfo{author}{{Van Grieken}, R.}, \bibinfo{author}{Markowicz, A.},
  \bibinfo{year}{2001}.
\newblock \bibinfo{title}{Handbook of X-Ray Spectrometry}.
\newblock \bibinfo{publisher}{CRC Press}, \bibinfo{address}{New York}.
%Type = Article
\bibitem[{VanBommel et~al.(2023)VanBommel, Berger, Gellert, O'Connell-Cooper,
  McCraig, Thompson, Fedo, {Des Marais}, Fey, Yen, Clark, Treiman and
  Boyd}]{vanbommel_2023}
\bibinfo{author}{VanBommel, S.}, \bibinfo{author}{Berger, J.},
  \bibinfo{author}{Gellert, R.}, \bibinfo{author}{O'Connell-Cooper, C.},
  \bibinfo{author}{McCraig, M.}, \bibinfo{author}{Thompson, L.},
  \bibinfo{author}{Fedo, C.}, \bibinfo{author}{{Des Marais}, D.},
  \bibinfo{author}{Fey, D.}, \bibinfo{author}{Yen, A.}, \bibinfo{author}{Clark,
  B.}, \bibinfo{author}{Treiman, A.}, \bibinfo{author}{Boyd, N.},
  \bibinfo{year}{2023}.
\newblock \bibinfo{title}{Elemental composition of manganese- and
  phosphorus-rich nodules in the {Knockfarril Hill} member, {Gale} crater,
  {Mars}}.
\newblock \bibinfo{journal}{Icarus} \bibinfo{volume}{392},
  \bibinfo{pages}{115372}.
\newblock \URLprefix
  \url{https://www.sciencedirect.com/science/article/pii/S001910352200464X},
  \DOIprefix\doi{https://doi.org/10.1016/j.icarus.2022.115372}.
%Type = Article
\bibitem[{VanBommel et~al.(2022)VanBommel, Gellert, Berger, McCraig,
  O'Connell-Cooper, Thompson, Yen, Boyd, Lanza and Ollila}]{vanbommel_2022}
\bibinfo{author}{VanBommel, S.}, \bibinfo{author}{Gellert, R.},
  \bibinfo{author}{Berger, J.}, \bibinfo{author}{McCraig, M.},
  \bibinfo{author}{O'Connell-Cooper, C.}, \bibinfo{author}{Thompson, L.},
  \bibinfo{author}{Yen, A.}, \bibinfo{author}{Boyd, N.},
  \bibinfo{author}{Lanza, N.}, \bibinfo{author}{Ollila, A.},
  \bibinfo{year}{2022}.
\newblock \bibinfo{title}{Constraining the chemical depth profile of a
  manganese-rich surface layer in {Gale} crater, {Mars}}.
\newblock \bibinfo{journal}{Spectrochimica Acta Part B: Atomic Spectroscopy}
  \bibinfo{volume}{191}, \bibinfo{pages}{106410}.
\newblock \URLprefix
  \url{https://www.sciencedirect.com/science/article/pii/S0584854722000544},
  \DOIprefix\doi{https://doi.org/10.1016/j.sab.2022.106410}.
%Type = Article
\bibitem[{VanBommel et~al.(2019a)VanBommel, Gellert, Berger, Yen and
  Boyd}]{vanbommel_2019B}
\bibinfo{author}{VanBommel, S.}, \bibinfo{author}{Gellert, R.},
  \bibinfo{author}{Berger, J.}, \bibinfo{author}{Yen, A.},
  \bibinfo{author}{Boyd, N.}, \bibinfo{year}{2019}a.
\newblock \bibinfo{title}{{Mars Science Laboratory} alpha particle {X-ray}
  spectrometer trace elements: Situational sensitivity to {Co, Ni, Cu, Zn, Ga,
  Ge, and Br}}.
\newblock \bibinfo{journal}{Acta Astronautica} \bibinfo{volume}{165},
  \bibinfo{pages}{32--42}.
\newblock \URLprefix
  \url{https://www.sciencedirect.com/science/article/pii/S0094576519312445},
  \DOIprefix\doi{https://doi.org/10.1016/j.actaastro.2019.08.026}.
%Type = Article
\bibitem[{VanBommel et~al.(2019b)VanBommel, Gellert, Boyd and
  Hanania}]{vanbommel_2019}
\bibinfo{author}{VanBommel, S.}, \bibinfo{author}{Gellert, R.},
  \bibinfo{author}{Boyd, N.}, \bibinfo{author}{Hanania, J.},
  \bibinfo{year}{2019}b.
\newblock \bibinfo{title}{Empirical simulations for further characterization of
  the {Mars Science Laboratory} alpha particle {X-ray} spectrometer: An
  introduction to the {ACES} program}.
\newblock \bibinfo{journal}{Nuclear Instruments and Methods in Physics Research
  Section B: Beam Interactions with Materials and Atoms} \bibinfo{volume}{441},
  \bibinfo{pages}{79--87}.
\newblock \URLprefix
  \url{https://www.sciencedirect.com/science/article/pii/S0168583X18307626},
  \DOIprefix\doi{https://doi.org/10.1016/j.nimb.2018.12.040}.
%Type = Unpublished
\bibitem[{VanBommel et~al.(2024)VanBommel, Sharma, Kizovski, Heirwegh,
  Christian, Knight, Ganly, Allwood, Hurowitz, Tice, Cable, Elam, Jones, Clark,
  Schmidt, Liu and Das}]{vanbommel_2024}
\bibinfo{author}{VanBommel, S.J.}, \bibinfo{author}{Sharma, S.},
  \bibinfo{author}{Kizovski, T.V.}, \bibinfo{author}{Heirwegh, C.M.},
  \bibinfo{author}{Christian, J.R.}, \bibinfo{author}{Knight, A.L.},
  \bibinfo{author}{Ganly, B.}, \bibinfo{author}{Allwood, A.C.},
  \bibinfo{author}{Hurowitz, J.A.}, \bibinfo{author}{Tice, M.M.},
  \bibinfo{author}{Cable, M.L.}, \bibinfo{author}{Elam, W.T.},
  \bibinfo{author}{Jones, M.W.M.}, \bibinfo{author}{Clark, B.C.},
  \bibinfo{author}{Schmidt, M.E.}, \bibinfo{author}{Liu, Y.},
  \bibinfo{author}{Das, A.}, \bibinfo{year}{2024}.
\newblock \bibinfo{title}{Constraining the cerium content of rocks and
  regolithin {Jezero} crater, {Mars}}.
\newblock \bibinfo{note}{Manuscript submitted for publication}.
%Type = Article
\bibitem[{Verma et~al.(2023)Verma, Maimone, Gaines, Francis, Estlin, Kuhn,
  Rabideau, Chien, McHenry, Graser, Rankin and Thiel}]{verma_2023}
\bibinfo{author}{Verma, V.}, \bibinfo{author}{Maimone, M.W.},
  \bibinfo{author}{Gaines, D.M.}, \bibinfo{author}{Francis, R.},
  \bibinfo{author}{Estlin, T.A.}, \bibinfo{author}{Kuhn, S.R.},
  \bibinfo{author}{Rabideau, G.R.}, \bibinfo{author}{Chien, S.A.},
  \bibinfo{author}{McHenry, M.M.}, \bibinfo{author}{Graser, E.J.},
  \bibinfo{author}{Rankin, A.L.}, \bibinfo{author}{Thiel, E.R.},
  \bibinfo{year}{2023}.
\newblock \bibinfo{title}{Autonomous robotics is driving {Perseverance}
  rover’s progress on {Mars}}.
\newblock \bibinfo{journal}{Science Robotics} \bibinfo{volume}{8},
  \bibinfo{pages}{eadi3099}.
\newblock \URLprefix
  \url{https://www.science.org/doi/abs/10.1126/scirobotics.adi3099},
  \DOIprefix\doi{10.1126/scirobotics.adi3099},
  \href{http://arxiv.org/abs/https://www.science.org/doi/pdf/10.1126/scirobotics.adi3099}{\tt
  arXiv:https://www.science.org/doi/pdf/10.1126/scirobotics.adi3099}.
%Type = Article
\bibitem[{Webster and Piccoli(2015)}]{webster_2015}
\bibinfo{author}{Webster, J.D.}, \bibinfo{author}{Piccoli, P.M.},
  \bibinfo{year}{2015}.
\newblock \bibinfo{title}{{Magmatic Apatite: A Powerful, Yet Deceptive,
  Mineral}}.
\newblock \bibinfo{journal}{Elements} \bibinfo{volume}{11},
  \bibinfo{pages}{177--182}.
\newblock \URLprefix \url{https://doi.org/10.2113/gselements.11.3.177},
  \DOIprefix\doi{10.2113/gselements.11.3.177},
  \href{http://arxiv.org/abs/https://pubs.geoscienceworld.org/msa/elements/article-pdf/11/3/177/3110886/177\_ELEM\_v11n3.pdf}{\tt
  arXiv:https://pubs.geoscienceworld.org/msa/elements/article-pdf/11/3/177/3110886/177\_ELEM\_v11n3.pdf}.
%Type = Article
\bibitem[{Wettergreen et~al.(2014)Wettergreen, Foil, Furlong and
  Thompson}]{wettergreen2014}
\bibinfo{author}{Wettergreen, D.}, \bibinfo{author}{Foil, G.},
  \bibinfo{author}{Furlong, M.}, \bibinfo{author}{Thompson, D.R.},
  \bibinfo{year}{2014}.
\newblock \bibinfo{title}{Science autonomy for rover subsurface exploration of
  the {Atacama} desert}.
\newblock \bibinfo{journal}{AI Magazine} \bibinfo{volume}{35},
  \bibinfo{pages}{47--60}.

\end{thebibliography}

\end{document}